\documentclass[sigconf]{acmart}

\AtBeginDocument{%
  \providecommand\BibTeX{{%
    \normalfont B\kern-0.5em{\scshape i\kern-0.25em b}\kern-0.8em\TeX}}}

\copyrightyear{2023}
\acmYear{2023}
\setcopyright{acmlicensed}\acmConference[CHI '23]{Proceedings of the 2023
CHI Conference on Human Factors in Computing Systems}{April 23--28,
2023}{Hamburg, Germany}
\acmBooktitle{Proceedings of the 2023 CHI Conference on Human Factors in
Computing Systems (CHI '23), April 23--28, 2023, Hamburg, Germany}
\acmPrice{15.00}
\acmDOI{10.1145/3544548.3580884}
\acmISBN{978-1-4503-9421-5/23/04}

\usepackage{gensymb}
\usepackage{multirow}
\usepackage{makecell}
\usepackage{appendix}

\begin{document}

\title[Exploring Different Levels of Autonomy and Machine Forms of Guiding Robots for the Visually Impaired]{"I am the follower, also the boss": Exploring Different Levels of Autonomy and Machine Forms of Guiding Robots for the Visually Impaired}

\author{Yan Zhang}
\orcid{0000-0003-2142-5094}
\affiliation{%
  \institution{Institute for AI Industry Research, Tsinghua University}
  \city{Beijing}
  \country{China}
}
\email{zhangyan@air.tsinghua.edu.cn}
\author{Ziang Li}

\orcid{0000-0003-3822-9145}
\affiliation{%
  \institution{Institute for AI Industry Research, Tsinghua University}
  \city{Beijing}
  \country{China}
}
\email{lza85995@gmail.com}

\author{Haole Guo}
\orcid{0000-0002-3453-6554}
\affiliation{%
  \institution{Institute for AI Industry Research, Tsinghua University}
  \city{Beijing}
  \country{China}
}
\email{guohaole@air.tsinghua.edu.cn}

\author{Luyao Wang}
\orcid{0000-0002-1786-3342}
\affiliation{%
  \institution{Institute for AI Industry Research, Tsinghua University}
  \city{Beijing}
  \country{China}
}
\email{wanglydesign@163.com}

\author{Qihe Chen}
\orcid{0000-0002-7133-2585}
\affiliation{%
  \institution{Institute for AI Industry Research, Tsinghua University}
  \city{Beijing}
  \country{China}
}
\email{cqh9933@163.com}

\author{Wenjie Jiang}
\orcid{0000-0002-6236-8623}
\affiliation{%
  \institution{Institute for AI Industry Research, Tsinghua University}
  \city{Beijing}
  \country{China}
}
\email{wenjie97strive@163.com}

\author{Mingming Fan}
\orcid{0000-0002-0356-4712}
\affiliation{%
  \institution{The Hong Kong University of Science and Technology (Guangzhou)}
  \city{Guangzhou}
  \country{China}
}
\affiliation{%
  \institution{The Hong Kong University of Science and Technology}
  \city{Hong Kong SAR}
  \country{China}
}
\email{mingmingfan@ust.hk}

\author{Guyue Zhou}

\orcid{0000-0002-3894-9858}
\affiliation{%
  \institution{Institute for AI Industry Research, Tsinghua University}
  \city{Beijing}
  \country{China}
}
\email{zhouguyue@air.tsinghua.edu.cn}

\author{Jiangtao Gong}
\authornote{Corresponding author}
\orcid{0000-0002-4310-1894}
\affiliation{%
  \institution{Institute for AI Industry Research, Tsinghua University}
  \city{Beijing}
  \country{China}
}
\email{gongjiangtao2@gmail.com}

\renewcommand{\shortauthors}{Yan Zhang, et al.}

\begin{abstract}
Guiding robots, in the form of canes or cars, have recently been explored to assist blind and low vision (BLV) people. Such robots can provide full or partial autonomy when guiding. However, the pros and cons of different forms and autonomy for guiding robots remain unknown. We sought to fill this gap. We designed autonomy-switchable guiding robotic cane and car. We conducted a controlled lab-study (N=12) and a field study (N=9) on BLV. Results showed that full autonomy received better walking performance and subjective ratings in the controlled study, whereas participants used more partial autonomy in the natural environment as demanding more control. Besides, the car robot has demonstrated abilities to provide a higher sense of safety and navigation efficiency compared with the cane robot. Our findings offered empirical evidence about how the BLV community perceived different machine forms and autonomy, which can inform the design of assistive robots.
\end{abstract}



\begin{teaserfigure}
  \centering{\includegraphics[width=1\textwidth]{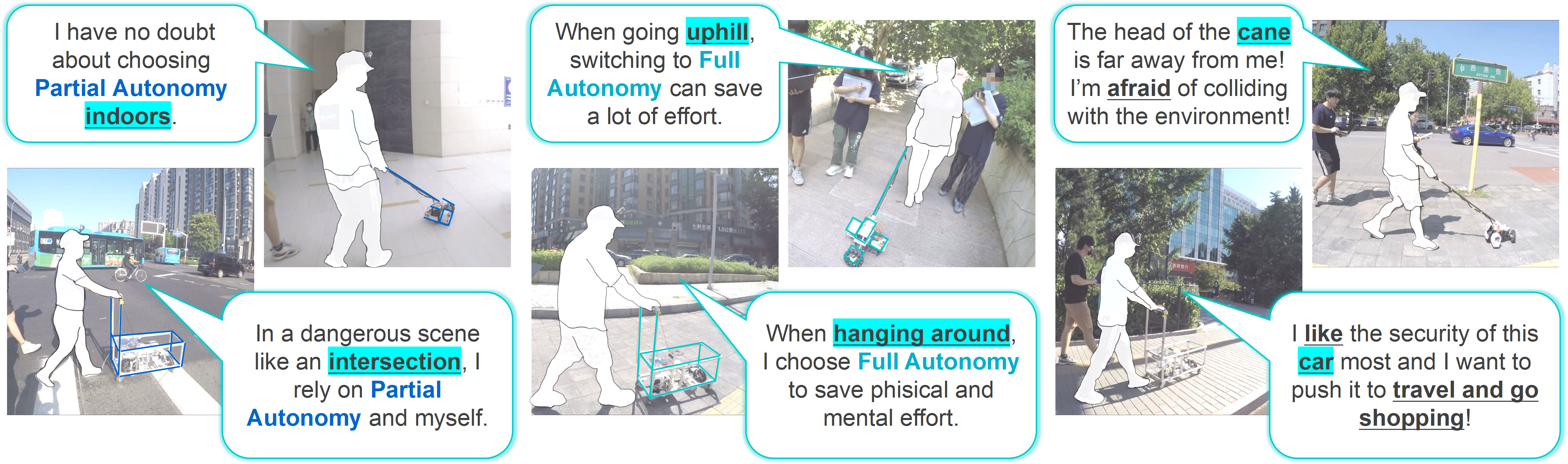}}
  \caption{An overview of participants' perception and utilization of different levels of autonomy and machine forms by demonstrating photos and interviews from the field study.}
  \label{fig:teaser}
\end{teaserfigure}

\maketitle

\section{Introduction}

Efficient and safe navigation assistive technologies have long been discussed as a solution to blind and low vision (BLV) groups' independent mobility difficulty since at least 2.2 billion people suffer from visual impairment \cite{who_2018}. The development of intelligent guiding devices has never stopped in the business and academic fields for a long time.

Up to now, various machine form designs have been tried~\cite{megalingam2019autonomous, wakita2011human, lee2013novel, xiao2021robotic, azenkot2016enabling, yasuda2020autonomous}. For the most efficient type of machine form with grounded kinesthetic feedback~\cite{slade2021multimodal}, there are two kinds, cane~\cite{slade2021multimodal, maidenbaum2014eyecane, wahab2011smart, takizawa2012kinect} and car~\cite{guerreiro2019cabot, tobita2018structure, xiao2021robotic, chen2021fuzzy}, that are primarily focused on by researchers. 
The machine form of interactive and assistive robots has an essential impact on the efficiency and experience of human-robot collaboration, such as trust \cite{adamik2021difference, esterwood2021you}, companionship \cite{bradwell2021morphology}, and communication \cite{kwon2010humanoid, rae2013influence}.
However, despite many assumptions and attempts that have been made, there is still a lack of thorough and detailed research on how the machine form of grounded guiding robot would affect the navigation efficiency and user experience. 

Meanwhile, different levels of autonomy are applied to the current guiding robots. A fully automated suitcase robot~\cite{guerreiro2019cabot} was invented for accurate navigation, which only required users to hold and follow, while a partially automated guiding cane~\cite{slade2021multimodal} was developed to turn and avoid obstacles, letting users control the forward motion by pushing. 
Autonomy level, one of the fundamental factors of human-robot interaction \cite{stubbs2007autonomy}, attracted many Human-Computer Interaction researchers to understand users' perception and utilization of levels of autonomy under different circumstances \cite{chanseau2016charge, nahavandi2017trusted}.   
However, whether partial autonomy is necessary for designing a guiding robot, which is valuable for understanding the BLV community and designing user-friendly guiding robots, has not been explored yet. 

To fill the above two gaps, in this paper, two guiding robots in the form of a cane and a car were created, and a controlled study and a field study with nineteen participants were accomplished to answer the questions above. 
The result showed that, contrary to popular belief, BLV people preferred the car other than the cane form robot, which may provide more sense of safety regardless of its inferiority in portability and familiarity. For the different levels of autonomy, participants showed various choices in different conditions. Full autonomy provided the highest enhancement in mobility and the best experience in a controlled and safe condition, whereas partial autonomy was the most preferable in a natural environment.

The contributions of this research are: i) two autonomy switchable guiding robots of cane and car form are designed and developed; ii) a controlled study to compare navigation efficiency and user experience of guiding robots with different levels of autonomy and machine forms; iii) a field study to investigate BLV's preference for autonomy level and machine form in a natural environment; iv) insights on guiding assistive robot design, including machine form and level of autonomy, respectively.

\section{Related Works}
In this section, we reviewed various grounded navigation robots focusing on the feature of machine form, followed by a deep digging into the human and assistive robot system in the aspect of autonomy level.

\subsection{Currently Grounded Robots with Different Machine Forms}
Compared to guiding devices using visual, audio, haptic, or non-grounded kinesthetic feedback \cite{lee2016rgb, lee2014wearable, tapu2020wearable, lin2012wearable, yang2021lightguide, ishihara2017beacon, bai2017smart, tapu2013smartphone, sato2019navcog3, ahmetovic2018turn, maidenbaum2014eyecane, wahab2011smart, takizawa2012kinect, sanchez2010autonomous, dastider2017cost,gong2020helicoach,ding2019helicoach}, guiding robots with grounded kinesthetic feedback that provides intuitive steering feedback has been proven to be more efficient in navigation \cite{slade2021multimodal}. The design framework for guiding robots is based on two main forms, the cane form, which is aligned with the perception and habits of visually impaired people, and the car form, which accommodates more on the stability of the robot mechanism.

\subsubsection{Cane Form Guiding Robots}
Guiding robotic canes have advantages like portability, lightweight, and familiarity to the BLV group. 
The philosophy of portable white canes and guide dogs, which provides a biological nature of kinesthetic feedback through the harness, was combined on a robotic cane "roji" with motor and wheels to provide navigation for visually impaired people \cite{shim2002robotic}. 
Suzuki, Hirata, and Ksuge \cite{suzuki2009development} developed a prototype of an intelligent passive cane with a similar robot structure but no active power that could guide users to the destination and avoid falling. Authors claimed cane form's advantages on the small size and lightweight. 
Another kind of guiding cane was mounted with a grounded rolling tip \cite{ye2016co, zhang2019human}. Ye et al. \cite{ye2016co} stated that their design was based on the fact that white canes have been widely used as mobility tools in the visually impaired community.

Nevertheless, constrained by the cane structure, cane form robots have significant shortcomings in stability, interference resistance, and load-bearing capacity. 
A user study with 10 BLV participants was conducted using a robotic cane that could navigate by tracing trails \cite{chuang2018deep}. The result indicated users' satisfaction with real-time guidance, reliable feedback, and a friendly interface. However, one trial failed as the participant pulled too hard, which prevented the robot from moving correctly. It reflected that, due to the mechanical structure, cane-form robots might not be stable and robust enough for uncontrollable environments and human behavior. 
Slade, Tambe, and Kochendefer (2021) designed a portable augmented cane with one steering omni-wheel to lead the turning direction \cite{slade2021multimodal}. Positive results of the mobility metrics demonstrated the design usability. Although authors applied various methods to reduce the weight, participants considered it heavy as the cane robot with one wheel was lack of self-balancing.

\subsubsection{Car Form Guiding Robots}
Besides the cane form, car-form robots are more common and compatible with robotics. Even though both have developed practical guiding functions, no comparison has been made to evaluate each form factor's impact on navigation efficiency and user experience. Carts, suitcases, or even quadruped robots can be categorized as car-form navigation assistive robots, which are physically conspicuous with superior in terms of integrated functionality, weight-bearing, and providing safety. 

The suitcase-shaped guiding robot, Cabot \cite{guerreiro2019cabot}, was designed to blend into the environment and mimic the interaction ability of guide dogs by locating on the left-hand side of users and equipped with an armrest. Cabot received high praise from BLV users on the feeling of comfort, confidence, and safety. 
Kayukawa et al. \cite{kayukawa2020blindpilot} proposed another car-form navigation robot that walked on the right side of users, which had consistency in selecting sensors and processors with the former robot. They assembled a LiDAR, stereo camera, and laptop, which could hardly be implemented on a white cane.
Besides being on the side, car-form guiding robots can be in front of users like a cart, such as LIGHBOT \cite{tobita2018structure}. LIGHBOT was developed to solve mobility problems inside facilities, which received positive feedback on easier, safer, and more confident movement.
Taking advantage of the structural stability of the car form, some guiding robots also assumed the function of walking assistance, which bore the dead weight as well as shared the weight of the human body leaning over \cite{lacey2000context, morris2003robotic}.

Furthermore, as the apparent distinctions between cane and car can bring essential impacts on users' cognition, while previous studies emphasized developing and improving one particular form attempting to disguise the weaknesses and amplify the strengths, BLV people's trade-off of two forms of guiding robots has not been studied in comparison. The overall perception brought to BLV by the superiority and limitation of cane and car form can be instructive for designing user-friendly guiding robots.

\subsection{Levels of Autonomy in Assistive Robots}
The autonomy level of navigation robots is evolving from semi-automatic to fully automatic, along with autonomous driving. 
The autonomy framework for robots to complete a task was divided into three aspects, sensing the environment, planning the movement, and acting upon the environment \cite{beer2014toward}. The allocation of control has classified the degree of autonomy in human-robot interaction: on humans, on robots, or shared control \cite{amirshirzad2019human, bauer2008human, stubbs2007autonomy}. 
To what aspect and extent robots should perform automatically is critical in achieving effective human-robot collaboration. 

Discussions have been made on human perceptions of different autonomy levels, such as sense of control \cite{chanseau2016charge, zafari2020attitudes}, trust \cite{nahavandi2017trusted}, and safety \cite{akalin2019evaluating}. 
The impact of partial and full autonomy of assistive robots for disabilities has been discussed in many aspects. There were findings indicating that, regardless of the high efficiency the full autonomy could provide, users needed access to control. 
However, there is a lack of discovery on visually impaired people's perception of partially and fully automated guiding robots and their needs and preference for changing autonomy levels in different situations.

A study researching disabled people's perception of autonomy level was based on a robot-assisted feeding system \cite{bhattacharjee2020more}. They proposed three types of autonomy that could provide users with various degrees of control. 
They found consistent results with the previous study that full autonomy could reduce users' effort and provide independence. However, participants commented that the existence of error refuted the conclusion that higher autonomy was always better. They also indicated a desire for an easy and direct way to control the speed. In another study \cite{kim2011autonomy}, the result of a three-week study on a robotic arm with both manual and autonomous modes, which was designed to assist subjects with traumatic spinal cords to complete pick-and-place tasks, showed that despite the auto-mode having reduced users' effort, their satisfaction was not raised as expected. 
Moreover, participants expressed their demand for having an input interface when using the autonomous mode, which reflected their desire to have control.

Various studies have researched different levels of autonomy or even autonomy-switchable walking assistive robots.
A study on safe walking technology figured that people who require walking assistive devices still need to take control \cite{holbo2013safe}.
Morris et al. \cite{morris2003robotic} proposed a car-form walking assistant robot for the elderly with both partial and full autonomy with a force sensor interface. The usability of autonomy-switchable walking robots has been proven. This design provided users with more flexibility to choose their desired level of autonomy but did not observe their preferences and performance in different modes.

As for the non-fully automated guiding assistive robot for visually impaired people, the allocation of control can be crucial as BLV lost 90\% of sensory information, which contains the essential clue for navigation. Even though there were surveys on the BLV group's attitude and desire to take control of an automated vehicle \cite{brinkley2020exploring, brewer2018understanding}, there was no empirical research providing insight into the actual behavior of BLV people towards mobility assistive robots with different autonomy and degree of control.
There has been some exploration of partially automated guiding robots.
Aigner and McCarragher \cite{aigner1999shared} developed a shared control robotic cane that could steer to avoid obstacles but left users to decide the route and destination. The design of the Augmented Cane \cite{slade2021multimodal} had the same idea of separating the forward and steering control to the user or robot but also enabled the cane to navigate to a preset goal. 
 
However, the above literature on guiding robots focused only on the development and deployment of a level of automation and was devoid of some consideration of the essential issues. We filled the research gap in understanding BLV users' perception of different levels of autonomy and their utilization of autonomy-switchable guiding robots. To accomplish this purpose, we designed two autonomy-switchable guiding robots with the cane form and the car form as study prototypes.

\section{Robot Design}
\begin{figure*}
    \centering
    \includegraphics[width=\textwidth]{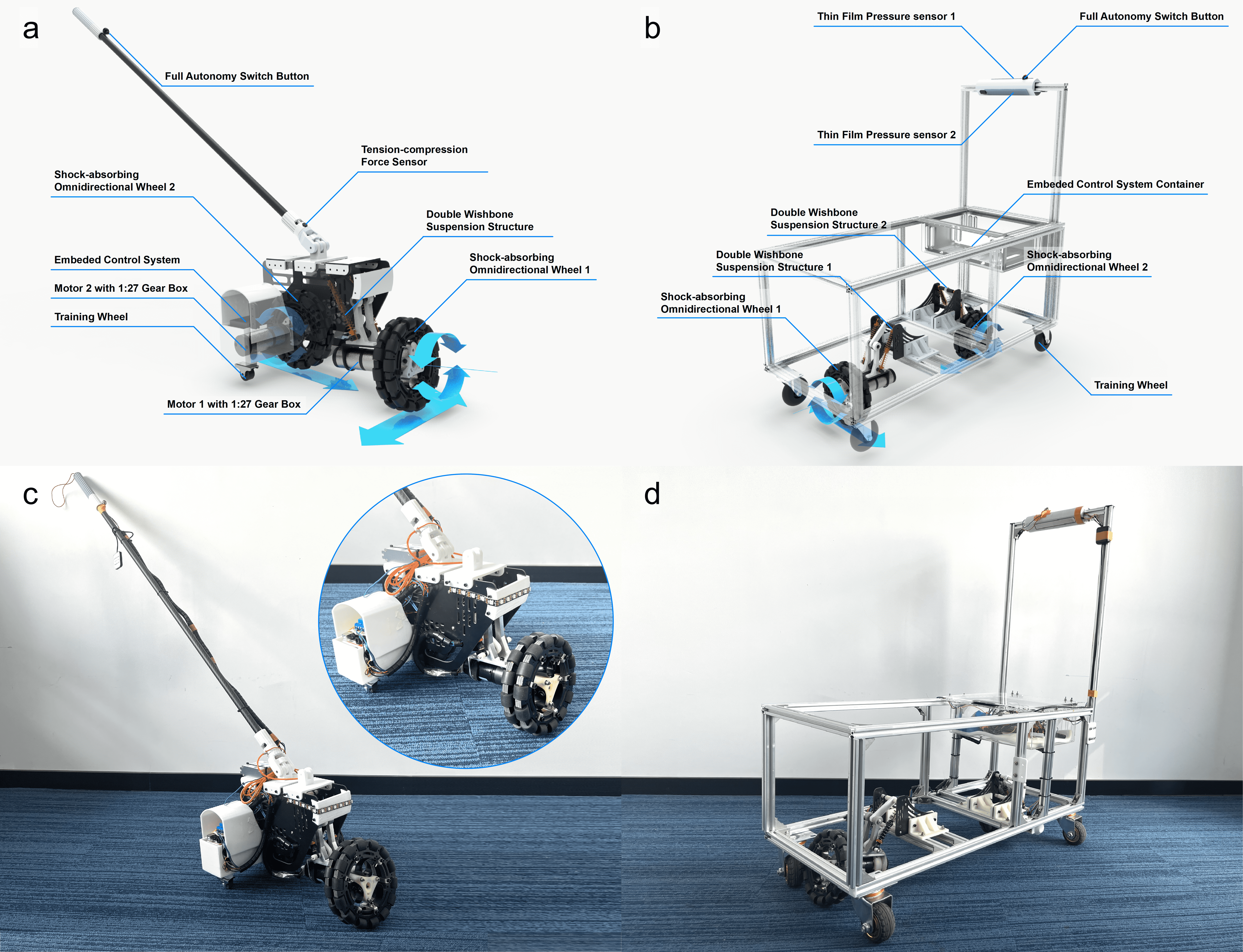}
    \caption{Models and photos of the final version of autonomy-switchable guiding robot cane and car. (a) The rendering model of the guiding cane. (b) The rendering model of the guiding car. (c) The photo of the guiding cane. (d) The photo of the guiding car. }
    \label{model}
\end{figure*}

To explore the BLV group's perception of the scale of the guiding robot, we offered both a cane and a car form (figure \ref{model}). Both of them had two different autonomous modes. BLV people controlled the speed of moving forward (or backward) by pushing (or pulling) using partial autonomy, and the robots guided the latitudinal steering. The counterpart, the full autonomy, provided both the movement in the longitudinal direction and turning guidance. Initially, we built the indoor version of those two guiding robots according to the controlled study environment. Nevertheless, considering the severe condition of road smoothness in the field study, we developed an upgraded version by adding a shock-absorbing structure and shared control interaction sensors which could let users switch autonomy by themselves.

\subsection{Hardware Design}
The cane and the car navigation robots shared most of their parts (see table \ref{table: hardware}). 

\begin{table*}[h]
\caption{Hardware Specification}
\label{table: hardware}
\resizebox{\textwidth}{44mm}{
\begin{tabular}{lcccc}
\hline
 & \multicolumn{2}{c}{Cane} & \multicolumn{2}{c}{Car} \\ \cline{2-5} 
 & Controlled Study & Field Study & Controlled Study & Field Study \\ \hline
Size & \makecell[c]{290mmx160mmx142mm,\\ with the cane length 1m} & \makecell[c]{347mmx334mmx199mm,\\ with the cane length 1m} & \multicolumn{2}{c}{770mmx360mmx300mm} \\ \hline
\makecell[l]{Structural\\ Material} & \makecell[c]{Fiberglass, carbon fiber,\\ and ABS resin} & \makecell[c]{Fiberglass, carbon fiber,\\ 6061 aluminum and ABS resin} & Square aluminum tube & \makecell[c]{Square aluminum tube,\\ 6061 aluminum, Acrylic sheet} \\ \hline
\makecell[l]{Motor, Gear,\\ and Encoder} & \makecell[c]{MD36NP27 (84 Watt, 325rpm)\\ with 1:27 gear ratio,\\ MD513 (4 Watt, 293rpm)\\ with 1:30 gear ratio,\\ and hall encoder} & \makecell[c]{MD36NP27 (84 Watt, 325rpm)\\ with 1:27 gear ratio,\\ and hall encoder} & \multicolumn{2}{c}{Square aluminum tube} \\ \hline
Wheels & \makecell[c]{Diameter 60mm \& 127mm,\\ Omni wheels;\\ Diameter 22mm,\\ plastic training wheels} & \makecell[c]{147mm shock-absorbing\\ Omni wheels;\\ Diameter 25.4mm,\\ rubber training wheels} & \makecell[c]{Diameter 127mm,\\ Omni wheels;\\ Diameter 48mm,\\ plastic training wheels} & \makecell[c]{147mm shock-absorbing\\ Omni wheels;\\ Diameter 80mm,\\ rubber training wheels} \\ \hline
Battery & \multicolumn{2}{c}{\makecell[c]{DC 12V battery\\(2000 mAh)}} & \multicolumn{2}{c}{\makecell[c]{DC 12V battery\\(5500 mAh)}} \\ \hline
Bluetooth & \multicolumn{2}{c}{HC-04} & \multicolumn{2}{c}{HC-04} \\ \hline
\makecell[l]{Motor \\Driver} & \multicolumn{2}{c}{Waynestark motor driver} & \multicolumn{2}{c}{Waynestark motor driver} \\ \hline
\makecell[l]{Embedded \\Controller} & \multicolumn{2}{c}{Arduino mega 2560} & \multicolumn{2}{c}{Arduino mega 2560} \\ \hline
Localization & HTC VIVE & GPS & HTC VIVE & GPS \\ \hline
\end{tabular}
}
\end{table*}

\subsubsection{Chassis Design}
Locomotion was achieved by the vector summation of vertically aligned lateral and longitudinal omni-directional wheel movements. The robot could reach partial autonomy by blocking the current to the longitudinal wheel's motor. Apart from the two active wheels, two and four training wheels for the cane and car, respectively, to share the weight and maintain balance.

To increase the passing capacity in the field study, we added unique omni-directional wheels with shock-absorbing ability and double wishbone suspension structures. We redesigned the traditional omni-directional wheel hub to leave additional cushioning space and selected the 0.3mm thick manganese steel sheet as the shock-absorbing material. Three of those sheets were all bent into an "$\Omega$" shape installed with the same interval to realize the shock absorption. The double wishbone suspension structure we designed was similar to the one on the vehicle. We put two springs on the left and right sides of the active wheels, and two links were installed in the middle (figure \ref{model}). Moreover, the motors were connected to the links by certain aluminum parts. 

\subsubsection{Form Factor}    
The size of the robot cane was designed in accordance with the white cane, and the size of the robot car was referenced to the guide dog. Furthermore, the handrail of the car was a rigid connection structure designed referring to the guide harness, which can better transmit kinesthetic feedback.

To adjust different users' heights and various holding positions of the robot cane, the cane holding in the user's hand was able to rotate up and down around its tip. To realize the same purpose for the robot car, the installation position of the handrail was adjustable. 
    
\subsection{Control Logic}
Both guiding robots shared the same control logic in the same version. In the controlled study, We deployed a positioning system and made the robot navigate autonomously. However, considering the field study's complicated environment and safety issues, we developed a remote control program to realize the Wizard-of-Oz method. A detailed description of the control method will be shown in the Study Approach in the Controlled Study and Field Study sections.

\section{Study 1: Controlled Study}
To understand how the machine form and level of autonomy influence the navigation efficiency and user experience in a controlled environment, we have conducted a 2 (machine form: cane, car) * 2 (level of autonomy: partial, full autonomy) controlled user study using the robots presented in the design section. 

\subsection{Study Approach}
\subsubsection{Participants}
\begin{table*}[h]
\caption{Demographic Information of Controlled Study Participants}
\label{table:control_participant}
\resizebox{\textwidth}{22mm}{
\begin{tabular}{ccccccccc}
\hline
PID & Gender & \makecell[c]{Age\\(years)} & Congenital/acquired & Level of BLV & \makecell[c]{Duration of BLV\\(years)} & \makecell[c]{Travel Frequency\\(/week)} & Travel aids & \makecell[c]{Full autonomy speed choice\\(Cane, Car)} \\ \hline
1 & Male & 25 & Congenital & Blind & 25 & 5 & White cane & 0.8, 0.8 \\
2 & Male & 33 & Acquired & Blind & 10 & 2 & White cane & 0.8, 0.8 \\
3 & Female & 36 & Acquired & Low vision & 10 & Hardly ever & White cane & 0.6, 0.6 \\
4 & Female & 33 & Congenital & Blind & 33 & 2 & White cane & 0.4, 0.4 \\
5 & Female & 29 & Acquired & Low vision & 10 & 14 & W/WO white cane & 0.4, 0.6 \\
6 & Male & 33 & Acquired & Blind & 3 & 7 & Guided by other people & 0.8, 0.6 \\
7 & Male & 47 & Congenital & Blind & 47 & 7 & White cane & 0.6, 0.6 \\
8 & Female & 38 & Congenital & Blind & 38 & 14 & Guided by other people & 0.8, 0.8 \\
9 & Male & 21 & Congenital & Blind & 21 & 3 & Guided by other people & 0.8, 0.8 \\
10 & Male & 29 & Congenital & Blind & 29 & 5 & White cane & 0.8, 0.8 \\
11 & Female & 34 & Congenital & Blind & 34 & 1 & White cane & 0.6, 0.8 \\
12 & Female & 31 & Congenital & Blind & 31 & 10 & White cane & 0.6, 0.8 \\ \hline
\end{tabular}
}
\end{table*}

Twelve visually impaired (male = 6 and female = 6) participants were recruited for this experiment. All participants had no other disability, and the age ranged from 21 to 47 (M=32.42, SD=6.54). Since our navigation robot system was designed to walk on the right-hand side of a person, subjects who were left-handed were screened out. The ability to use white canes was one of the screening criteria. A detailed demographic of visually impaired participants are listed in table \ref{table:control_participant}. We collected the assistive travel method (travel aids) commonly used by each participant. Most of them rely on the white cane to travel, and three always have other people to guide them. One participant with low vision sometimes would not use the white cane and walked by her feelings. 
All participants with varying degrees of visual impairment were blindfolded to control extraneous variables. With the approval of the Institutional Review Board (IRB), we have received written informed consent from all participants.

\subsubsection{Apparatus and System}
\begin{figure*}[h]
  \centering
  \includegraphics[width=0.8\linewidth]{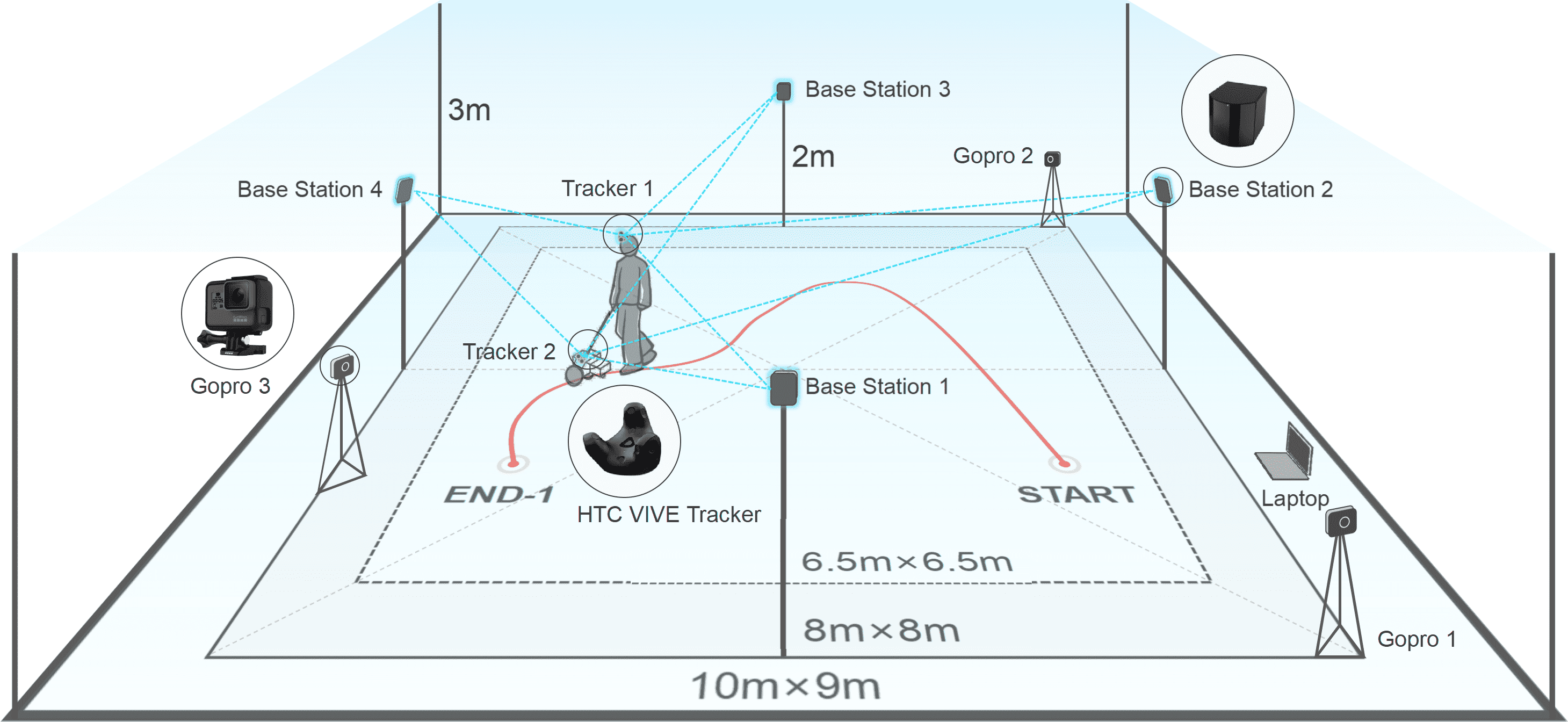}
  \caption{The experiment setup schematic of the controlled study. The guiding cane and the path on the ground are drawn as demonstrations.}
  \label{figure:control_setup}
\end{figure*}

Figure \ref{figure:control_setup} shows that the controlled experiment was deployed in a 10m*9m hall, enclosed to avoid interruptions. Both two forms of robots led BLV people to follow pre-set routes in an 8m*8m square area. The apparatus, which are the robots, carried one HTC VIVE tracker at the front and an Arduino UNO with an HC-04 Bluetooth module to receive locomotion instructions. And the system consisted of four HTC VIVE base stations located at the midpoint of each side of the square area and a server laptop calculating and sending locomotion instructions. (The details about the trail-following method can be found in the appendix \ref{Controlled study trail-following method}) Three Gopros were installed at different angles for video recording. There were randomly placed obstacles outside the range of the route to make the simulation more realistic. The control commands of motors were updated at a frequency of 4Hz.  

To eliminate the influence of individual differences on mobility, a traditional white cane and a normal cart, which has the same scale as the car robot, were used as the baseline.
Each participant used the white cane, the cane robot with partial and full autonomy, the normal cart with the same scale as the car robot, and the car robot with partial and full autonomy during the experiment. 

We offered three handrail styles for the car robot, the suitcase-type, the cart-type, and the harness-type like the guide dog. Participants could decide the handrails' most suitable installation style and height for themselves.  
Before the experiment begins, participants can select a walking speed for the full autonomy robots from 0.4m/s, 0.6m/s, and 0.8m/s. Participants' selections are listed in table \ref{table:control_participant}.
Participants were guided by the audio instruction displayed in the Bluetooth headphones when using the non-powered cane and cart. The PC server calculated positions and played audio automatically. When the participant walked on the left side of the route, it would play "Right." When the participant was on the right side of the route, "Left" was played. When using partial autonomy, participants could push the robot with a freely forward speed and turn with the steering generated by the robot. When using full autonomy, participants could follow the robot that has a pre-selected velocity.

A self-design phone app was used to assist with the secondary task, which will be introduced in the Task section. The app can randomly select one of two vibration modes every 3 seconds to perform. While using the robot with their right hand, participants were instructed to hold the phone with their left hand and touch the screen if the current vibration mode was the same as the last one. The app then recorded the number of times the screen was touched and calculated the accuracy.

\subsubsection{Task}
The experiment had two tasks with parallel relationships: the main walking task and the secondary memory task. 

The main task was aimed at examining participants' walking performance. The route of the walking task consists of six different turns (45\degree left turn, 90\degree left turn, 135\degree left turn, 45\degree right turn, 90\degree right turn, and 135\degree right turn) in random arrangements and straight walks in between. Similar route design principles can be seen in \cite{yang2021lightguide, liu2021tactile, ahmetovic2018turn, xu2020virtual}.
Due to the size of the site, the entire route was designed to be a round-trip; each was 10 meters long and contained three different turns. Subjects completed the route in two runs for each trial with a 2-minute interval. 

The secondary task was aimed at examining participants' mental workload by using the N-back task. A 1-back task with two vibration modes was performed throughout the walking task. The 1-second short and 3-second long vibrations appeared randomly every 3 seconds, and subjects had to judge within the interval to click or not to click the screen. Clicking the screen indicated they believed the current vibration mode was the same as the previous one; otherwise, it indicated inconsistency.
A similar technique for measuring BLV people's mental effort while using a navigation assistive device has been used in \cite{zhao2020effectiveness}.

\subsubsection{Procedure}
Before the experiment, there was a training phase to familiarize participants with the robots and tasks. The training routes were also randomly generated. Participants used each assistive equipment to navigate training routes until they claimed they had mastered it. Moreover, experimenters demonstrated three forward speeds of the full autonomy mode for participants and asked them to choose the one that best suited them. The chosen speed was implemented during the experiment. Subsequently, participants attempted the 1-back task until they were deemed to have understood the rule, achieving an 80\% accuracy in a 30s trial.

The sequence of experimental conditions was counterbalanced. When participants claimed they were ready, the program randomly determined a route task at the beginning of each trial. Every participant departed from the same point (figure \ref{figure:control_setup}). The secondary task started and finished simultaneously with the main task. The system recorded the real-time speed, off-course distance, duration, human path length, and the accuracy of the 1-back task. After each trial, participants filled in a Psychosocial Impact of Assistive Devices Scale (PIADS) \cite{day2002development} and System Usability Scale (SUS) \cite{brooke1996usability}. 
After the experiment ended, there was a structured interview, and the questions are presented in appendix \ref{control_interview}. The entire process was video recorded.

\subsubsection{Evaluation}
\begin{enumerate}
\item Walking performance: 
To analyze how the guiding robot affected participants' walking efficiency, the walking performance metrics included speed, duration, path length, and off-course distance. 
To eliminate the bias of individual mobility, each variable was subtracted from the baseline without autonomy (normal white cane and cart). 

\item Mental workload: 
To measure participants' mental effort when collaborating with guiding robots, the accuracy of the secondary 1-back task was calculated and analyzed.

\item Usability:
Each robot's usability can be revealed by SUS results (see appendix \ref{appendix: scales}).

\item Psychosocial impact:
PIADS was designed to evaluate the impact of assistive devices on functional independence, quality of life, and well-being. As measuring the effect of assistive robots on trust and satisfaction with the sense of control was part of the experiment purposes, these two items were added to PIADS as a supplement (see appendix \ref{appendix: scales}). 

\item Subjective feeling:
The post-test interview was conducted to receive participants' subjective feelings.

\end{enumerate}
The two-way repeated-measures analysis of variance (ANOVA) \cite{st1989analysis} was conducted based on marginal means to analyze quantitative measurement. Post-hoc tests were conducted to identify pairwise comparisons. The normal distribution of variables and homogeneity of variances were examined in advance. Data that did not conform to a normal distribution were analyzed by Friedman's two-way ANOVA test \cite{friedman1937use}. The chosen confidence interval was 95\%. 

\subsection{Results and Discussion}
\begin{figure*}[h]
  \centering
  \includegraphics[width=\textwidth]{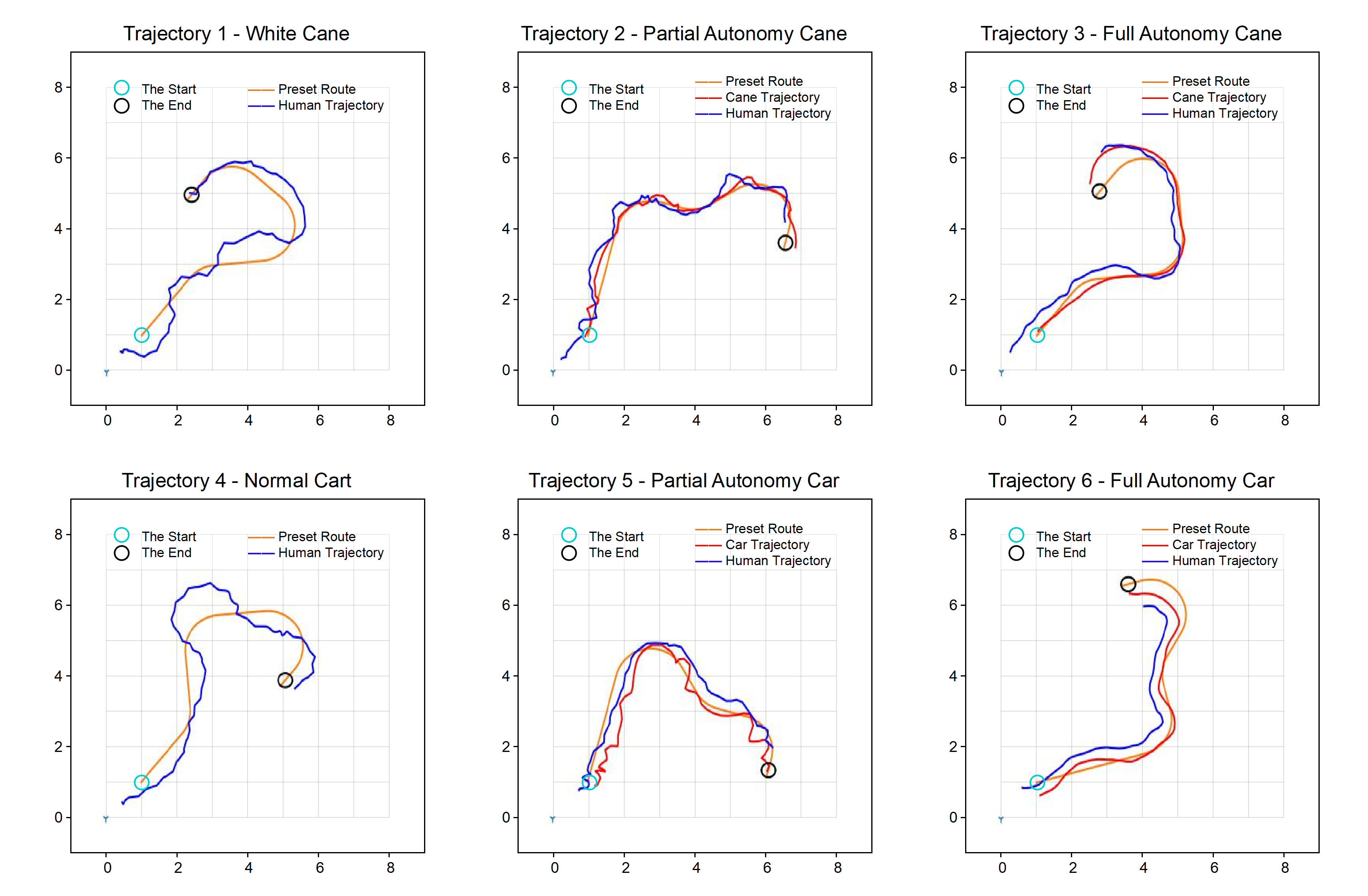}
  \caption{Examples of the preset route, robot trajectory, and human trajectory in the controlled study.}
  \label{figure:control_trajectory}
\end{figure*}

All participants have successfully completed the experiment, and 24 trials of data have been analyzed. Illustrations of the experiment trajectory results of each guiding method are shown in figure \ref{figure:control_trajectory}. The control program randomly generated the preset route (orange) before the navigation began. Human (blue) and robots' (red) trajectories were recorded and plotted by trackers. It can be seen that the user's path matched the preset route more when using guiding robots than the baseline white cane and cart.

\begin{figure*}[h]
  \centering
  \includegraphics[width=0.8\linewidth]{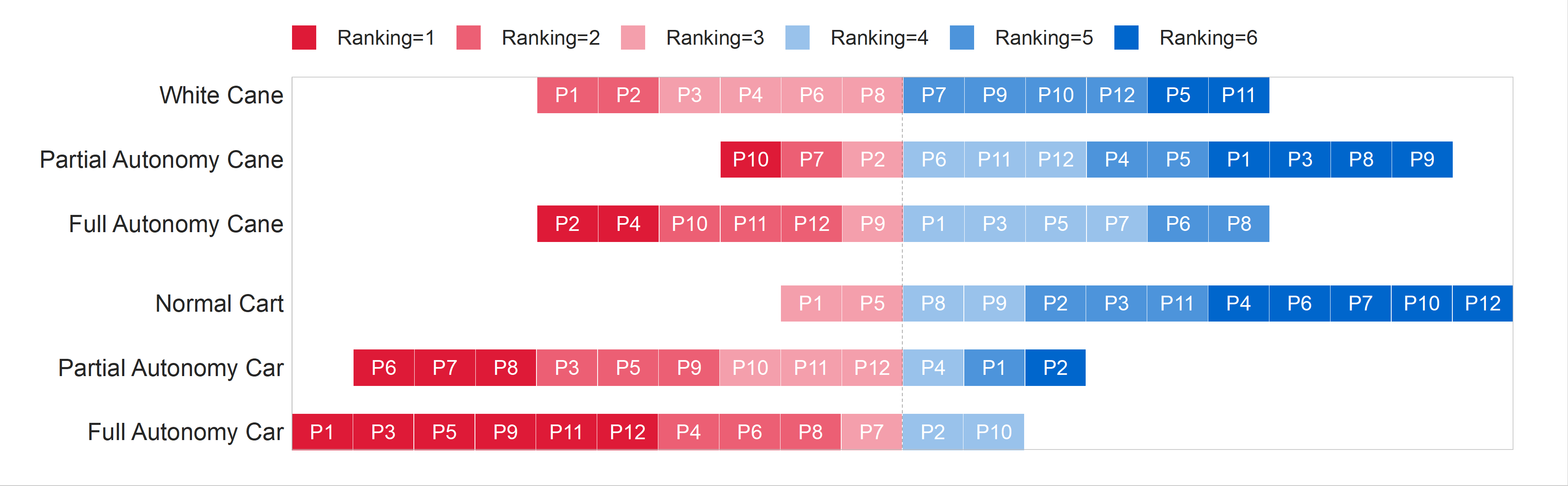}
  \caption{Participants' preference ranking of the baseline white cane, baseline normal cart, and robots with different forms and autonomy levels.}
  \label{figure:preference}
\end{figure*}

The overall preference ranking of each combination of autonomy and machine form is shown in figure \ref{figure:preference}. In summary, the fully automated guiding car was the most favored, followed by the partially automated car, fully automated cane, and partially automated cane. Even though one participant expressed enjoyment of partial autonomy cane, 4 out of 12 BLV listed it in the last place. Participants had a neutral attitude towards the traditional white cane; the normal cart was the least popular.        

\subsubsection{Navigation Efficiency}
\begin{table}[]
    \caption{Results of Walking Performance Metrics (Mean (Standard Deviation)).}
    \label{table:control}
    \resizebox{0.47\textwidth}{16mm}{
    \begin{tabular}{ccccc}
    \toprule
      & \multicolumn{2}{c}{\makecell[c]{Partial Autonomy\\(Bench-marked)}} & \multicolumn{2}{c}{\makecell[c]{Full Autonomy\\(Bench-marked)}} \\
    \cline{2-5} 
     & Cane & Car & Cane & Car \\
     \midrule
    Speed (m/s) & -0.04 (0.17) & 0.14 (0.18) & -0.02 (0.12) & 0.08 (0.12) \\
    Duration (s) & 3.64 (6.77) & -3.00 (4.18) & -3.36 (5.64) & -4.19 (2.21) \\
    Path Length (m) & 1.02 (1.90) & 0.19 (1.92) & -0.34 (2.06) & -2.12 (2.33)\\
    Off-course Distance (m) & -0.12 (24.75) & -6.19 (15.60) & 0.52 (17.99) & -33.77 (16.96) \\
    \bottomrule
    \end{tabular}
    }
\end{table}

\begin{figure*}
    \centering
    \includegraphics[width=0.8\textwidth]{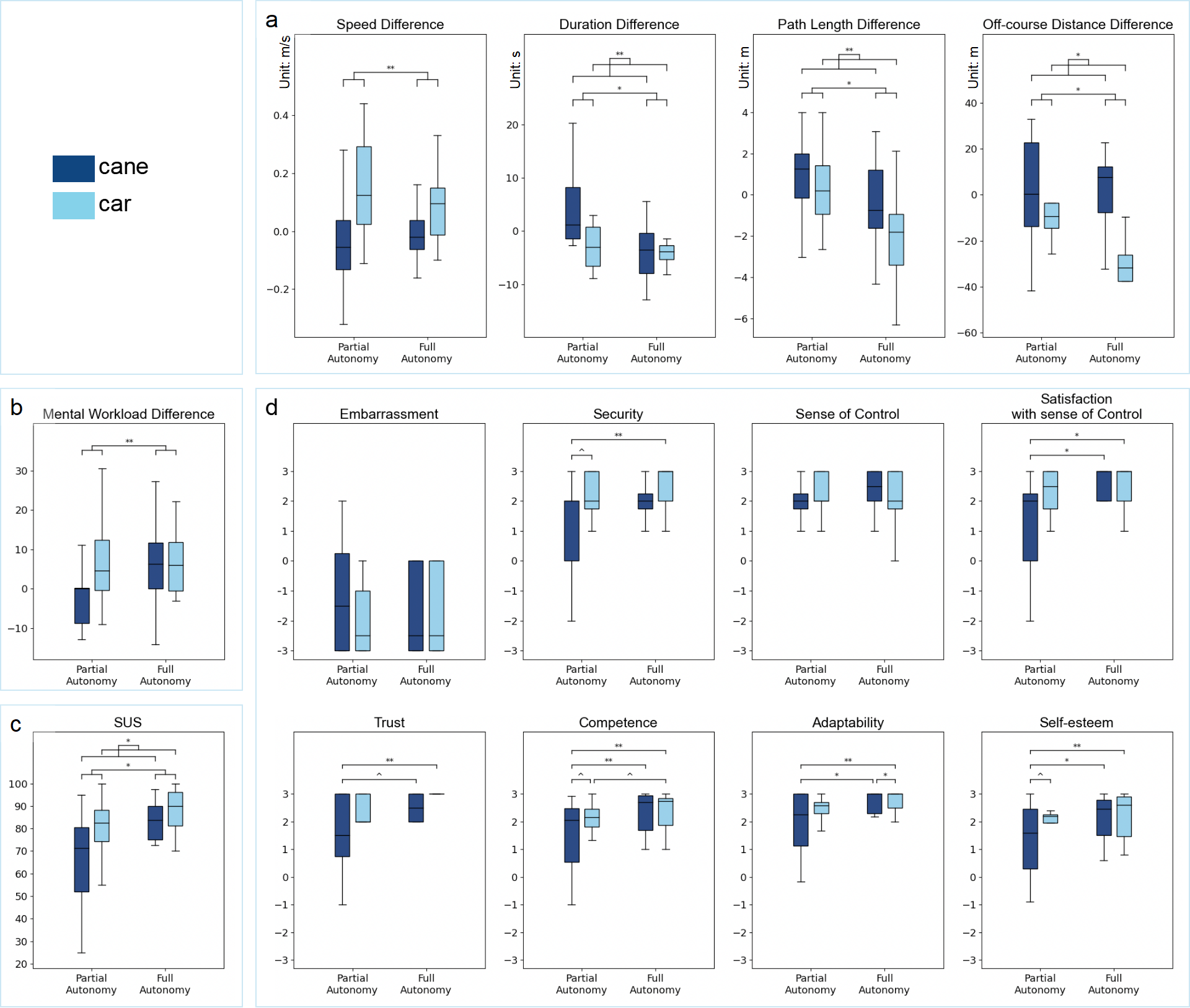}
    \caption{Controlled study result box-plots. (a) The result of navigation efficiency. (b) The result of mental workload measurement. (c) The result of SUS. (d) The result of PIADS. }
    \label{figure:comb_box}    
\end{figure*}

As mentioned in the Evaluation section, the walking performance metrics have been uniformly bench-marked to offset individual differences. The results are shown in table \ref{table:control} and figure \ref{figure:comb_box}. 

As for the walking speed, participants tended to have a higher speed when using the guiding car. A significance has been found in autonomy. Participants had a higher speed when using full autonomy ($M=0.05, SD=0.19$) than partial autonomy ($M=0.03, SD=0.13, F=9.93, p=0.01$). No evidence has been found in machine form and interaction. 

In terms of duration, every robot surpassed the baseline except the partial autonomy cane ($M=3.64, SD=6.77$). An interaction effect has been found between the autonomy level and machine form ($F=5.71, p=0.04$). The post-doc analysis indicated that machine form only had an effect when using partial autonomy. The partially automated car ($M=-3.00, SD=4.18, F=10.73, p=0.01$) had a shorter duration than the partially automated cane ($M=3.64, SD=6.77$). 
Moreover, the autonomy level only had an effect when using a guiding cane, as the fully automated guiding cane ($M=-3.36, SD=5.64, F=9.19, p=0.01$) had a better result than the partially automated car ($M=3.64, S=6.77$). It was significantly shorter when navigated by car ($M=-3.60, SD=3.32$) than cane ($M=0.14, SD=7.06, F=4.94, p=0.05$) and by full autonomy ($M=-3.78, SD=4.21$) than partial autonomy ($M=0.32, SD=-6.46, F=9.37, p=0.01$). 

For the trajectory length, only the guiding car reduced the distance traveled, which indicated BLV could follow the car more precisely. Full autonomy ($M=-1.23, SD=2.33, F=4.69, p=0.05$) had a shorter path length than partial autonomy ($M=0.60, SD=1.92$). The result of the car-form robot ($M=-0.97, SD=2.40, F=17.02, p<0.01$) was also significantly shorter than the cane-form robot ($M=0.34, SD=2.06$). However, no interaction effect was found in the aspect of path length. 

The off-course distance was the sum of the lateral deviation distance from the route. It was significantly smaller when navigating with the car-form robot ($M=-16.09, SD=22.29, F=16.78, p=0.03$) than the cane-form robot ($M=2.00, SD=19.47$). The off-course distance was also shorter when using full autonomy ($M=-13.52, SD=24.76, F=5.74, p=0.04$) than partial autonomy ($M=-1.02, S=19.14$). A tendency of interaction effect ($F=4.21, p=0.07$) can be seen, which has been confirmed by post-hoc that the machine form had an effect under full autonomy that the full autonomy car form robot ($M=-33.7, SD=16.93, F=20.98, p<0.01$) had a significantly better result than the full autonomy cane robot ($M=0.52, SD=17.99$). The autonomy level had an effect with a car-form robot that the fully automated robotic car ($M=-33.77, SD=16.93, F=9.90, p=0.01$) had a better result than the partially automated robotic cane ($M=-6.19, SD=15.60$).

The walking performance metrics consistently demonstrated that the car-form and full autonomy robot could provide better guidance and resulted in superior mobility compared with a normal cane and cart. 

\subsubsection{User Experience}
\begin{table*}[h]
\caption{Results of User Experience Metrics}
\label{table:experience}
\begin{tabular}{cccccccc}
\hline
 &  & Mental workload & SUS & \multicolumn{4}{c}{PIADS} \\ \cline{3-8} 
 &  & \makecell[c]{Accuracy\\(Bench-marked)} & Score & Competence & Adaptability & Self-esteem & Mean \\ \hline
\multirow{2}{*}{\makecell[c]{Partial\\Autonomy}} & Cane & -3.07\% (10.03) & 65 (22.41) & 1.52 (1.25) & 1.92 (1.15) & 1.38 (1.27) & 1.61 (0.56) \\
 & Car & 7.85\% (11.96) & 80 (12.72) & 2.01 (1.29) & 2.36 (1.03) & 2.01 (1.06) & 2.13 (0.14) \\ \hline
\multirow{2}{*}{\makecell[c]{Full\\Autonomy}} & Cane & 5.32\% (13.73) & 81 (12.72) & 2.28 (0.84) & 2.51 (0.76) & 2.11 (0.88) & 2.3 (0.09) \\
 & Car & 5.54\% (11.35) & 88 (10.24) & 2.40 (0.85) & 2.58 (0.70) & 2.20 (0.84) & 2.40 (0.08) \\ \hline
\end{tabular}
\end{table*}
Participants' accuracy of the 1-back task was calculated for the mental workload. Benchmarking has been performed to reduce the effects of individual differences in memory. The results were all positive, which meant a reduction in the mental workload, except for the partially automated cane (table \ref{table:experience}). The results of two-way ANOVA presented a significant interaction effect. The post-hoc analysis proved that the machine form only had an effect under partial autonomy, as the partial autonomy guiding car ($M=7.85, SD=11.96, F=24.46, p<0.01$) had a better result than the partial autonomy cane ($M=-3.07, SD=10.03$).
The autonomy level had an effect when using the cane robot, as the fully automated cane ($M=5.54, SD=11.35, F=7.75, p=0.02$) had a better result than the partially automated cane ($M=-3.07, SD=10.=03$).
The accuracy was significantly larger when using full autonomy guiding robots ($M=5.43, SD=12.32, F=12.51, p=0.01$) than partial autonomy guiding robots ($M=2.39, SD=12.15$).

The SUS scores of partial autonomy cane, full autonomy cane, partial autonomy car, and full autonomy car were reported as 65, 81, 80, and 88, respectively (table \ref{table:experience}). The results of two-way ANOVA on SUS indicated participants' preference for full autonomy ($M=84.69, SD=11.89, F=16.97, p=0.02$) than partial autonomy ($M=72.60, SD=19.44$), and the car form ($M=84.27, SD=12.03, F=7.19, p=0.02$) than the cane form ($M=73.02, SD=19.62$). No significant interaction effect has been found (figure \ref{figure:comb_box}).

The PIADS revealed users' satisfaction with assistive guiding robots in multiple aspects. Friedman's two-way ANOVA test has evaluated ratings on different machine forms and levels of autonomy. Figure \ref{figure:comb_box} shows the distribution and statistical differences of five items and three indicators. 

For the measure of embarrassment, the partial autonomy car ($M=-2.00, SD=1.21$) could reduce participants' embarrassment the most, followed by the full autonomy car ($M=-1.75, SD=1.42$) and the full autonomy cane ($M=-1.67, SD=1.50$), while no significance was found. Regarding safety, significance has been found among different robots ($\chi^2 = 13.87, p<0.01$). The full autonomy car showed the highest safety ($M=2.42, SD=0.79$), significantly higher than the partial autonomy cane ($M=1.25, SD=1.48, p<0.01$). The partial autonomy car ($M=2.08, SD=0.79$) also demonstrated the same tendency to surpass the partial autonomy car ($p=0.06$). The trust item had a similar pattern to safety, in which full autonomy car ($M=2.83, SD=0.39, p=<0.01$) and full autonomy cane ($M=2.50, SD=0.52, P=0.08$) were rated higher than partial autonomy cane ($M=1.58, SD=1.44$). No evidence for the dissimilarity in the sense of control has been found, but satisfaction with the sense of control had a different result ($\chi^2=8.53, p=0.04$). Full autonomy cane received the highest satisfaction with the sense of control ($M=2.42, SD=0.90$), followed by full autonomy car ($M=2.33, SD=0.98$). Both of them had a significantly higher score than the partial autonomy cane ($p=0.05, p=0.02$). 

The results of three indicators and mean PIADS scores of different forms and autonomy combinations are demonstrated in table \ref{table:experience} and figure \ref{figure:comb_box}. Full autonomy cane and car have been considered the most helpful for increasing the feeling of independence, well-being, and quality of life.

In conclusion, full autonomy and car-form robot provided the best user experience comprehensively in the controlled study.

\subsubsection{Interview}
During the interview, most participants expressed their preference for full autonomy and the car-form robot. They reported car forms' advantages, such as can provide a higher sense of safety, can block more obstacles, and having spaces to integrate more functions. They also shared their concerns on guiding cars, including requiring more practice because of unfamiliarity compared to a cane and difficulty in getting through narrow passages. The benefits of the cane robot have also been mentioned; for example, it is quick to learn and can be adapted to more situations. However, low safety was the biggest issue. 

For different levels of autonomy, participants complimented full autonomy in many aspects, such as it is worry-free and effortless, can be comfortably handled, and has high safety. As convenient as full autonomy was, they felt partial autonomy was unnecessary and had little desire to be in control. However, they complained that full autonomy could not match their walking speed perfectly. On the contrary, partial autonomy can satisfy this demand and provide comfort and control, but it requires more effort. P8 explained her understanding of different autonomy and imagined the utilization in different circumstances:
\begin{quote}
"Full autonomy is quite necessary when going to an unfamiliar environment. Partial autonomy helps when I am not too familiar with the place, especially when I am unfamiliar with turning. In fact, there are times when BLV people are particularly eager to run, even if jogging. We crave a sense of that speed stimulation, so partial autonomy can help me when I'm running at my speed. I think each person's choice is due to their desire to be in control."
\end{quote}

\subsection{Summary}
As the cane and car form autonomy switchable guiding robots received superior SUS scores (81, 88) compared to previous studies (68.125 \cite{morris2003robotic}, 65 \cite{slade2021multimodal}, 88 \cite{guerreiro2019cabot}, and 77,1 \cite{tobita2018structure}), it is confident to say that both the guiding cane and car have acceptable usability. It was consistently revealed by quantitative and qualitative results of the controlled study that full autonomy car-form robots had advantages in guiding efficiency and user experience. For the autonomy level, participants had greater velocity, shorter duration, shorter path length, smaller off-course distance, and less mental workload when using full autonomy. Full autonomy provided a higher sense of safety, trust, and satisfaction with control. For the machine form, car-form robots reduced travel duration, path length, and off-course distance for visually impaired people, with a higher sense of safety and trust.

However, even though partial autonomy was ranked higher than white cane and normal cart in preference, it has been described by BLV as "unnecessary" and "helpful, but not as beneficial as full autonomy." We hypothesized that the control experiment did not restore their mental state in a real environment and that the impact factors demonstrating participants' perceptions of different autonomy levels and machine forms have not been fully explored. 

To further discover visually impaired people's preference for autonomy and machine form in daily use, we upgraded our robots and conducted a field study. 

\section{Study 2: Field Study}
Shortly after the first experiment, we conducted a field study with upgraded guiding robots. We aimed to figure out under what circumstance BLV people would use partial or full autonomy and would their preference for machine forms changes in real on-road situations. In this study, participants used both guiding robots to travel to their desired destination and were allowed to switch autonomy directly at their will. The navigation and obstacle avoidance functions were realized by the Wizard-of-Oz method.

\subsection{Study Approach}
\subsubsection{Preparation}
\begin{figure*}[h]
  \centering
  \includegraphics[width=0.8\textwidth]{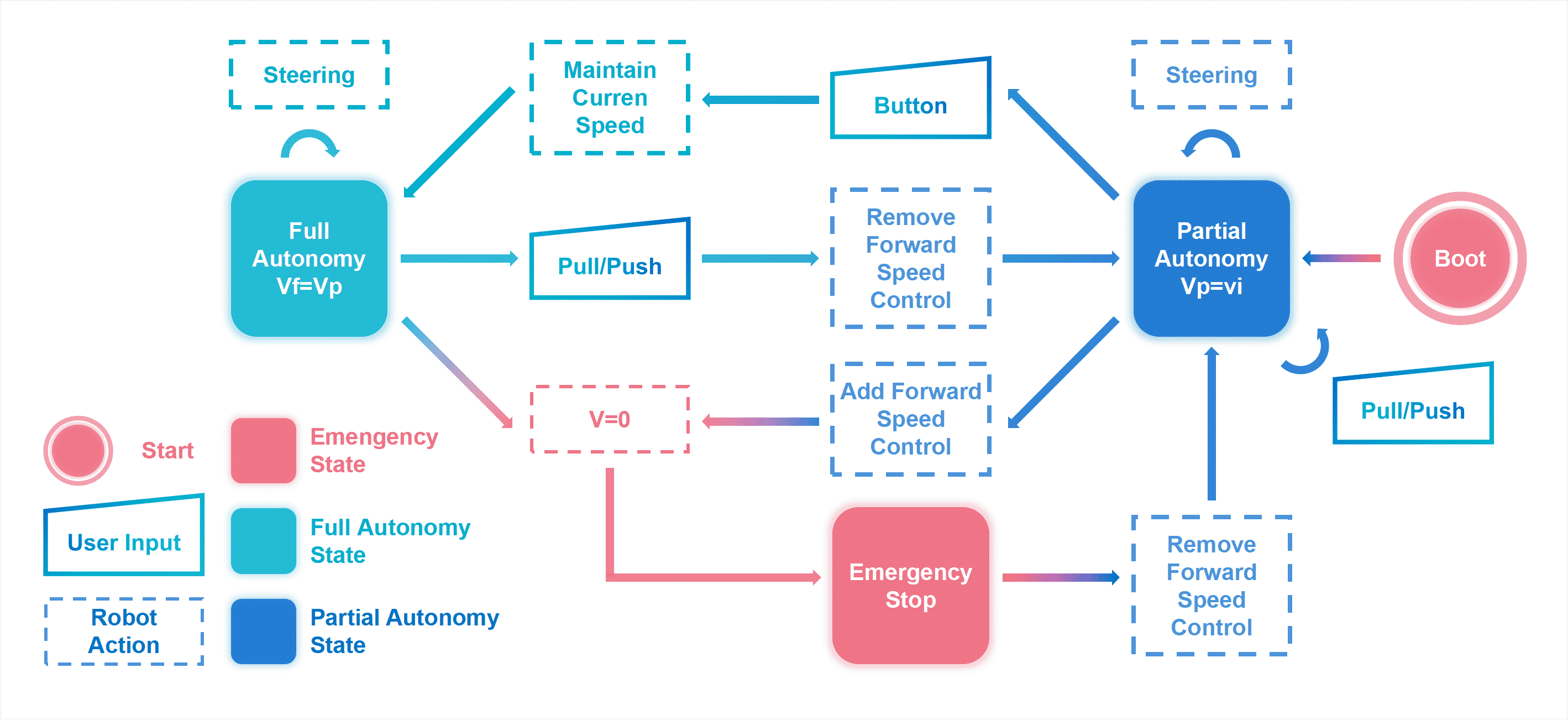}
  \caption{The finite-state machine of the upgraded autonomy-switchable robot cane and car in the field study.}
  \label{FSM}
\end{figure*}

The user interface and control logic has been redesigned to improve the robots' practicality and achieve users' active autonomy switch. A pilot field study with two participants, who completed a route with indoor and outdoor parts, was conducted to evaluate the usability. After collecting their comments and iterating the design, the final version is shown as figure \ref{model}. 
\begin{enumerate}
\item Finite-state machine:
A button and force sensors have been implemented on the handrail, enabling the microcontroller to learn about users' intent of changing autonomy. Figure \ref{FSM} illustrates the finite-state machine of the cane and car-guiding robots. After booting the robot, partial autonomy is set to the default state. Users control the forward speed ($V_{p}$) by following their desires by pulling and pushing the handrail. When pressing the button mounted on the handrail, the state will be switched to full autonomy, and the forward speed will be maintained at the current speed by the motor ($V_{f}=V_{p}$). If the force sensor perceives any pull or push action, indicating the user's intention to take over, the motor will yield the forward control and switch back to partial autonomy. Guiding robots share the lateral control to turn and avoid obstacles by steering at a partial autonomy state. For safety reasons, both states have an emergency stop protocol. Specifically speaking, if it is at partial autonomy, the forward motor will take over the longitudinal control when facing an unavoidable obstacle. The onward speed will be reduced to zero immediately ($V_{p}=0$). After eliminating the crisis, the control will be relocated to the user. If it is at full autonomy, the state will be switched to partial autonomy after breaking is completed, and the danger disappears.

\item Control logic: 
In the apparatus of this improved version, the HC-04 Bluetooth module, which was used on the apparatus in controlled study, was kept for wireless data transfer but the tracker was replaced by a GPS module connected to the upgraded controller: Arduino Mega. The system, which was quite simple compared with the one used in the controlled study, was just a laptop, and the input was acquired through the keyboard controlled by an operator. In addition to the basic forward speed and steering control, there were keys for emergency brake and autonomy switching. After testing, the operation delay of the remote control system was less than 100ms, which was enough to use the wizard-of-Oz method during this study.

\end{enumerate}

\subsubsection{Procedure}
Three massage parlors have been selected as the study sites because of the aggregation of BLV crowds and the diversity of neighborhoods. Before the field study, agreements were made with participants, including their choices of destinations. All destinations are within approximately 10 minutes of walking distance. Participants departed using one guiding robot and returned using the other. Orders were counterbalanced. Three participants with low vision wanted to be blindfolded during the study, as they claimed that they had the ability to travel under the sunlight discreetly but could never walk independently during the night. They expected to simulate the condition at night and were banking on the robots to improve their nighttime mobility.

Initially, participants all signed the informed consent release form, which was approved by the Institutional Review Board (IRB). A 10 minutes training section was scheduled before the experiment started, which ensured the understanding and mastery of the use of guiding robots. The target location was communicated in advance, and they had a general idea of the direction and path. During the field trip, participants were allowed to switch autonomy to their free will. The emergency avoidance protocol was followed as figure \ref{FSM} with experimenters' verbal notification to guarantee safety.  Because of the robot chassis's limitation, experimenters helped move the robot over when encountering stairs. Each study lasted one hour, including training and interview, and each participant received 100 CNY as a reward.

\begin{figure*}[h]
  \centering
  \includegraphics[width=0.8\linewidth]{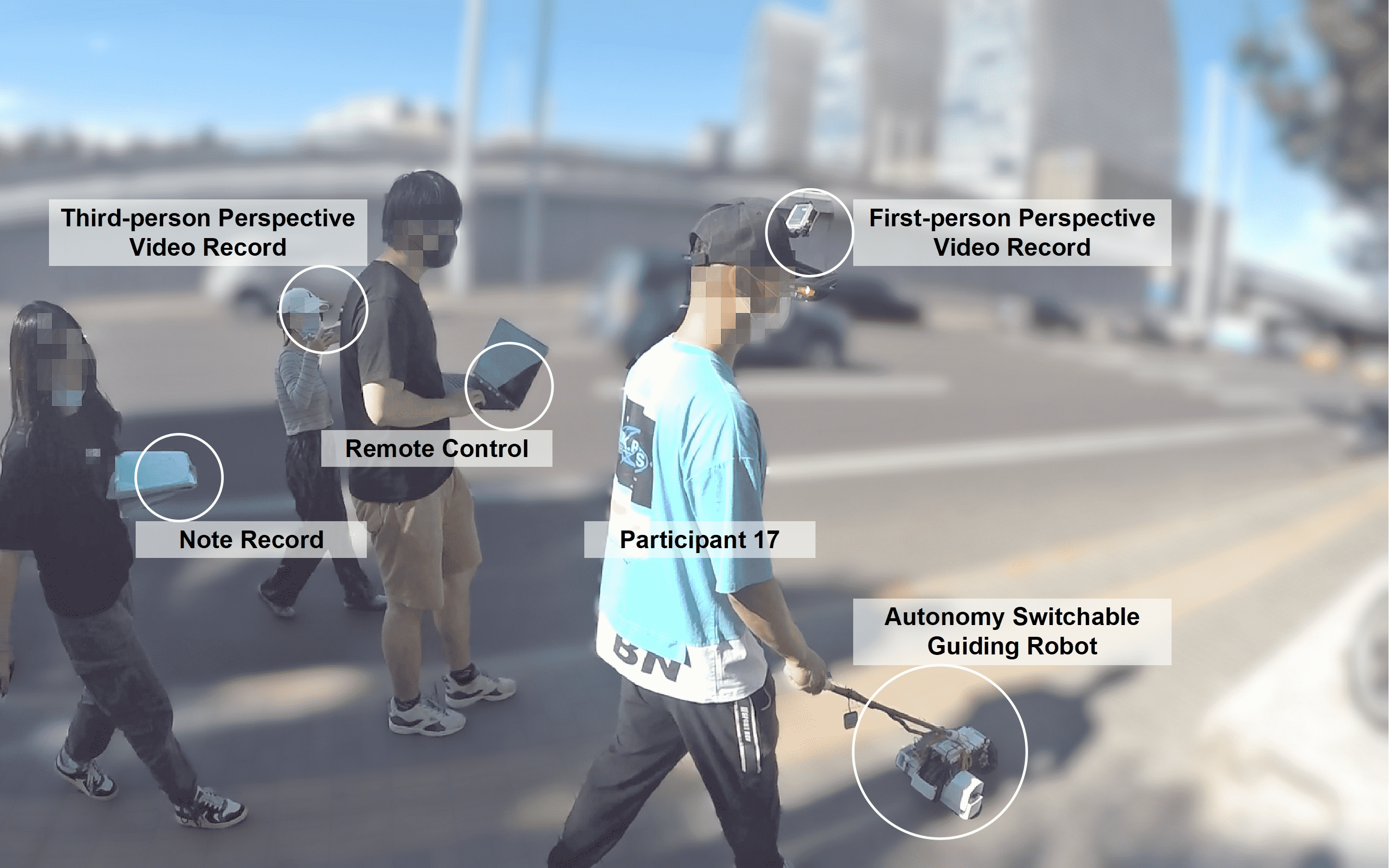}
  \caption{The on-site experiment setup illustration, showing participant 17 as an example.}
  \label{site}
\end{figure*}

There were four researchers on-site during the field study (figure \ref{site}), with one who was responsible for the remote control of the guiding robot in the Wizard-of-Oz style, one whom video recorded the entire process, and two who were in charge of observation and post-experiment interview. In addition to the third-person perspective recorded by the researcher, the first-person perspective was also recorded by the participants' head-mounted camera. All participants were encouraged to express real-time feelings and the reason for autonomy-switching actions during the walk. A post-experimental structured interview was conducted after each trial. The details of the interview questions are presented in Appendix \ref{field_interview}. Subjects' real-time speed and location were recorded by guiding robots.

\subsubsection{Participants}
\begin{table*}[t]
    \caption{Demographic Information of Field Study Participants ($^*$: Blindfolded)}
    \label{table:field_participant}
    \resizebox{\textwidth}{20mm}{
    \begin{tabular}{cccccccc}
    \toprule
    PID & Gender & \makecell[c]{Age\\(years)} & Congenital/acquired & Level of BLV & \makecell[c]{Duration of BLV\\(years)} & \makecell[c]{Travel Frequency\\(/week)} & Travel aids \\
    \midrule
    5$^*$ & Female & 29 & Acquired & Low vision* & 10 & 14 & W/WO white cane \\
    8 & Female & 38 & Congenital & Blind & 38 & 14 & Guided by other people \\
    13 & Male & 31 & Congenital & Blind & 11 & 2 & White cane \\
    14$^*$ & Male & 22 & Congenital & Low vision* & 22 & 14 & W/WO white cane \\
    15$^*$ & Female & 25 & Congenital & Low vision* & 25 & Hardly ever & Guided by other people \\
    16 & Male & 33 & Acquired & Low vision & 9 & 1 & White cane \\
    17 & Male & 34 & Acquired & Blind & 10+ & 4 & White cane \\
    18 & Male & 48 & Acquired & Blind & 10+ & Hardly ever & White cane \\
    19 & Male & 42 & Acquired & Blind & 8 & Hardly ever & White cane and Guided by other people \\
    \bottomrule
    \end{tabular}}
\end{table*}

Nine participants (3 females, 6 males) with ages ranging from 22 to 48 (M=33.56, SD=8.17) completed the field study, which included two congenital blindness, three acquired blindness, and four low vision. Regarding mobility, seven of them can travel with or without a white cane independently. Table \ref{table:field_participant} itemizes participants' detailed demographic. All subjects worked at the three massage parlors and self-reported being familiar with the surrounding areas. P5 and P8 have been involved in both controlled and field studies.

\subsubsection{Evaluation}
\begin{figure*}[h]
  \centering
  \includegraphics[width=1\textwidth]{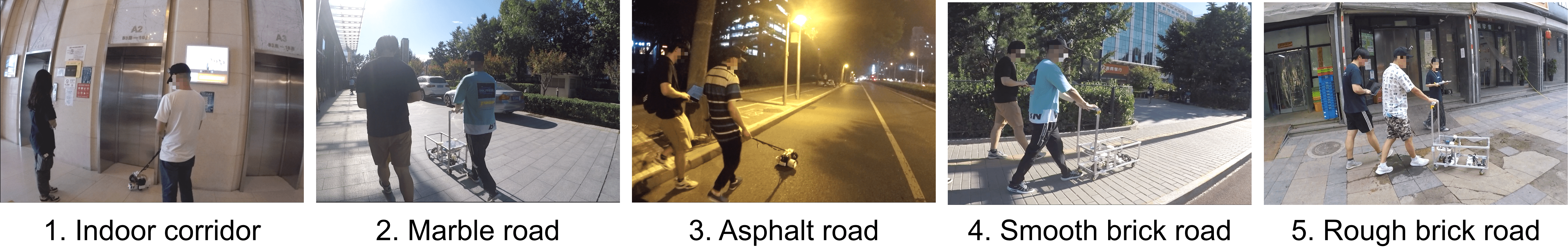}
  \caption{Examples of different road surface conditions.}
  \label{road_condition}
\end{figure*}

Videos have been manually coded by two researchers following the same coding scheme (Table \ref{coding_scheme}) to analyze the potential environmental impacts on users' autonomy choices. Road features, including surface smoothness, slope, intersection, width, moving objects, and environmental noise, have been defined. Figure \ref{road_condition} demonstrates the surface condition of each category. The indoor corridor was the most smooth one, followed by the marble road. Vibration could be felt through the handrail when walking on the asphalt road, but the feeling underfoot was still flat. The smooth brick road had more pronounced brick projections, and a slight undulation could be felt as the front wheel passed through the gaps between bricks. The rough brick road had the worst surface, with irregular cracks and the risk of tripping.

The counting method of the coding was based on the number of times each category appeared per autonomy section. The inter-rater Kappa of two researchers coded the video was greater than 0.9 (p<0.01). Paired t-tests \cite{kim2015t} and Wilcoxon tests \cite{cuzick1985wilcoxon} have been used to analyze each variable according to the normality. A two-way ANOVA test has been conducted on human walking speed and duration ratio of use for each autonomy.

\begin{table}[h]
    \caption{The Manual Coding Scheme of Video Recordings.}
    \label{coding_scheme}
   \resizebox{0.47\textwidth}{44mm}{
    \begin{tabular}{lll}
    \toprule
    Road Feature & Category & Description \\
    \midrule
    \multirow{5}{*}{Surface} & Indoor corridor & \multirow{5}{*}{\makecell[l]{Increasing roughness \\from indoor corridor to\\ rough brick road.}} \\ 
     & Marble road &  \\
     & Asphalt road &  \\
     & Smooth brick road &  \\
     & Rough brick road &  \\
    \midrule
    Slope & \makecell[l]{Uphill/Downhill\\/No slope} &  \\
    \midrule
    Intersection & Yes/No &  \\
    \midrule
    Width & Narrow/Wide & \makecell[l]{Need or no need to\\ give way to people \\walking in the \\opposite direction} \\
    \midrule
    \multirow{3}{*}{Moving objects} & Pedestrian & \multirow{3}{*}{\makecell[l]{The object was counted \\as it was moved\\ towards the participant.}} \\
     & Bicycle and motorcycle &  \\
     & Vehicle &  \\
    \midrule
    \multirow{3}{*}{\makecell[l]{Environmental\\ noise}} & Talking & \multirow{3}{*}{\makecell[l]{The type and quantity\\ were counted.}} \\
     & \makecell[l]{Non-motor Vehicle \\Sound} &  \\
     & Motor Vehicle Sound &  \\
     \bottomrule
    \end{tabular}
    }
\end{table}

\subsection{Observations and Findings}
In summary, 7738 seconds of video data have been collected from 18 trips performed by 9 participants. Everyone chose a different destination, and three of them experienced an indoor section. In general, participants navigated around their daily workplace using guide robots and encountered various on-road conditions, including intersections with and without traffic lights, mixed lanes for pedestrians and vehicles, up and downhills, wide and narrow sidewalks, etc.

For the overall preference for partial autonomy and full autonomy, both quantitative and qualitative data show that the former was superior. The duration analysis revealed a significantly higher utilization of partial autonomy ($M=0.71, SD=0.14$) than full autonomy ($M=0.29,  SD=0.14$) with $t=4.56, p=0.002$. Seven out of nine participants expressed their favor for partial autonomy during the interview. The other two who preferred full autonomy were P15 and P19. 

In terms of walking speed, the two-way ANOVA showed that users' average speed using full autonomy ($M=0.613, SD=0.028$) was significantly higher than partial autonomy ($M=0.524, SD=0.015$) with $F=12.836, p=0.012$, which indicates that participants tend to have the navigation robot with full autonomy help them maintain a faster speed. The variance of the real-time speed of partial autonomy ($M=0.026, SD=0.004$) was significantly larger than full autonomy ($M=0.0134, SD=0.002$) with $F=13.968, p=0.01$, which indicates participants tend to change their speed more substantially when using partial autonomy. P16 reported that he set a higher speed when using full autonomy because he would switch when noticing the environment was safe so that he could walk more efficiently and boldly. \textit{"I would like to try an even faster speed in a safe environment," said P16.}

\subsubsection{On-site Case Demonstration}
\begin{figure*}
  \centering
  \includegraphics[width=\textwidth]{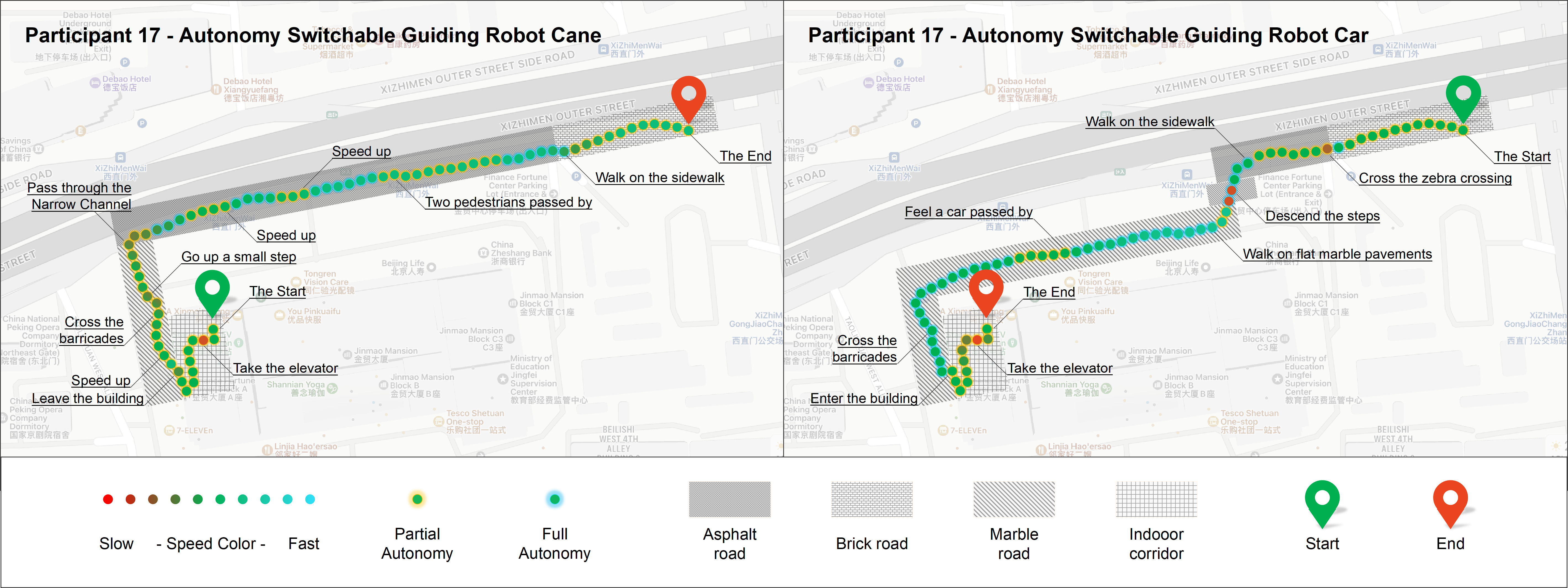}
  \caption{Examples of field Study trajectories (participant 17).}
  \label{field_trajectory}
\end{figure*}

Figure \ref{field_trajectory} provides a complete view of the trajectories of participant 17. He started from the indoor corridor using the guiding cane with partial autonomy. After leaving the building, he decided to speed up on the marble road. Before stepping on the pedestrian-vehicle mixed asphalt road, he encountered several narrow sections and stairs. During this section, he switched on and off full autonomy several times. The first time he exited full autonomy was for speed adjustment as he decided to walk faster and pushed the robot cane. The second time he felt two pedestrians pass by in front of him, he pulled the robot to stop even though the guiding cane was steering to avoid them. After getting on the brick sidewalk, he used partial autonomy until reaching the destination. The return travel was navigated by the robot car. He used partial autonomy before getting to the zebra crossing. As there was a car passing by, but the participants didn't slow down, the operator followed the emergency protocol to restrict the forward speed as well as provide a verbal notification. There was a short section on the asphalt road before the downward steps. After getting down the stairs, the participant walked on a long stretch of flat marble pavements, during which he chose to use full autonomy most of the time and only switched to partial autonomy when he felt a car pass by. While entering the building, he decelerated and changed to partial autonomy. Participant 17 returned to the start point after an elevator ride. 

\subsubsection{Scenario-related Factors}
The video coding results of the on-road conditions are shown in table \ref{coding_result}. 
In general, BLV people used more partial autonomy when sensing a complex, dangerous, and changeable context.

\begin{table}
    \caption{Results of Scenario Analysis.}
    \label{coding_result}
    \resizebox{0.47\textwidth}{34mm}{
    \begin{tabular}{lllll}
    \toprule
    Variable & Partial Autonomy & Full Autonomy & Value & p \\
    \midrule
    Indoor corridor & M=0.37 & M=0.11 & t=5.94 & $0.027^*$ \\
    Marble road & M=0.42 & M=0.36 & t=0.71 & 0.528 \\
    Asphalt road & M=0.60 & M=0.47 & t=2.39 & $0.044^*$ \\
    Smooth brick road & Mdn=0.42 & Mdn=0.40 & Z=-0.59 & 0.553 \\
    Rough brick road & Mdn=0.38 & Mdn=0.13 & Z=-2.20 & $0.028^*$ \\
    \midrule
    Uphill & M=0.11 & M=0.11 & t=-0.40 & 0.711 \\
    Downhill & Mdn=0.17 & Mdn=0.00 & Z=-1.75 & $0.08^\wedge$ \\
    \midrule
    Intersection & M=0.33 & M=0.15 & t=3.15 & $0.01^{**}$ \\
    \midrule
    Narrow section & M=0.49 & M=0.41 & t=1.69 & 0.129 \\
    Wide section & Mdn=0.83 & Mdn=1.00 & Z=1.58 & 0.114 \\
    \midrule
    Pedestrian & Mdn=1.92 & Mdn=1.60 & Z=-2.07 & $0.038^*$ \\
    Bicycle and motorcycle & M=1.49 & M=1.04 & t=2.00 & $0.08^\wedge$ \\
    Vehicle & Mdn=0.77 & Mdn=0.00 & Z=-2.67 & $0.008^{**}$ \\
    \midrule
    Talking & M=0.71 & M=0.49 & t=1.01 & 0.34 \\
    Non-motor vehicle sound & Mdn=0.40 & Mdn=0.25 & Z=-1.90 & $0.058^\wedge$ \\
    Motor vehicle sound & M=2.43 & M=1.45 & t=3.36 & $0.01^{**}$ \\
    \bottomrule
    \end{tabular}
    }
\end{table}

Regards the \textbf{(1) road surface condition}, participants chose to use partial autonomy significantly more times than full autonomy when walking on rough sections like rough brick roads ($Mdn_{partial}=0.38, Mdn_{full}=0.13, p=0.028$) and asphalt roads ($M_{partial}=0.60, M_{full}=0.47, p=0.044$). Although the indoor corridor had the smoothest surface, it was also narrow (86\% of the narrow section was the indoor corridor) and winding that partial autonomy was used more than full autonomy ($M_{partial}=0.37, M_{full}=0.11, p=0.027$). In contrast, the \textbf{(2) width} of the road section did not show a sign of a clear impact on BLV's choice of autonomy. 
During the interview, P14 claimed that he could feel the road's bumper from the handrail's vibration and the noise from the wheels of the guiding car.\textit{"The car shook more frequently,  and the wheels hit the bumps on the ground with a sound. ... I believed that the environment had become complicated, so I chose to push it myself with partial autonomy."} P17 explained the reason for using partial autonomy indoors, even if the floor was flat. \textit{"The interior space is not large, and often having to turn. I think it is better for me to control the speed."}

As for the \textbf{(3) slope}, the results showed a tendency that partial autonomy was more frequently chosen than full autonomy during the downhill ($Mdn_{partial}=0.17, Mdn_{full}=0.00, p=0.08$), whereas there was no significant difference when facing the uphill ($M_{partial}=0.11, M_{full}=0.11, p=0.553$). Nevertheless, participants clearly expressed their considerations when facing a slope. P8 and P16 had major concerns about speed control downhill. \textit{"It's handy to use partial autonomy to access the speed. ... I was worried that it would cause sudden impact when maintaining the speed going downhill"}, said P8. P8, 15, and 16 agreed that using full autonomy to go uphill can save more effort. However, P8 switched to partial autonomy midway up the hill as there was a turn in the wheelchair-accessible route. \textit{"I was unsure when I felt the steering, so I changed back to partial autonomy"}, said P8.

Moreover, the existence of \textbf{(4) intersections} was determined as a strong influence on users' choice as significantly more partial autonomy has been used than full autonomy when passing the crosswalk ($M_{partial}=0.33, M_{full}=0.15, p=0.01$). P13, 14 and 15 believed it would be safer to directly control the forward speed themselves in such varied and hazardous scenes. P14 explained that when he heard the sound of traffic nearby, he was afraid to walk. P5 additionally stated that she was worried about the guiding robot (in partial autonomy) pulling her forward, even though she knew it could safely avoid obstacles and stop at any time. Unlike the others, P19 chose to use full autonomy when crossing as he was more anxious about not getting through the intersection in time for the green light phase. \textit{"I was aware that I was slow on foot, and there was a time limit. It would be safer to set a higher forward speed using full autonomy and let it lead me to go through."}

For the \textbf{(5) moving objects}, the Wilcoxon signed-rank test revealed a significantly higher chance for participants to switch to partial autonomy when more pedestrians ($Mdn_{partial}=1.92, Mdn_{full}=1.60, p=0.038$) or vehicles ($Mdn_{partial}=0.77, Mdn_{full}=0.00, p=0.008$) were approaching or passing by. A slight trend can be found for the number of moving bicycles and motorcycles after running the t-test ($M_{partial}=1.49, M_{full}=1.04, p=0.08$). Respecting the \textbf{(6) environmental noise}, despite the number of talking voices ($M_{partial}=0.71, M_{full}=0.49, p=0.34$) and non-motor vehicle sounds ($Mdn_{partial}=0.40, Mdn_{full}=0.25, p=0.058$) didn't show significance, the higher density of motor vehicle sound resulted in significantly greater preference in partial autonomy ($M_{partial}=2.43, M_{full}=1.45, p=0.01$). All participants have directly or indirectly referenced their ability to feel objects around them, including seeing, hearing, and senses they cannot describe. Stated they would rely on partial autonomy when they believed the surroundings were complex as it can provide a higher sense of safety.

However, it is worth noticing that BLV's feelings are not always correct. There was a time when P17 suddenly took over the control when walking in a relatively quiet and open area. He self-reported that he felt a car was coming toward him, but there was not. 

\subsubsection{Machine Form Related Factors}
For the aspect of machine form, the most obvious difference is the \textbf{(1) size}. Figure \ref{model} and table \ref{table: hardware} illustrate the scale of each robot in detail.  Most participants mentioned the sense of safety when asked about the difference in experience between the two forms' sizes. They believed the physically large and sturdy car could help them block additional hazards. P13 and P15 specifically described the impact of length, width, and height of a guiding robot on sensation and usability:
\begin{quote}
P15: "My sense of safety actually depends on the width (of the robot). As for the length, as big as a suitcase will do. The handrail had better be folded so I could squeeze on the subway."\\
P14: "The size of the car could be slightly smaller. It's in the right width, but the length can be reduced, as it's the width that makes me feel protected."
\end{quote} 

\hspace*{\fill} \

Apart from the difference in size, there were other variables found that affected participants' experience. Firstly, the \textbf{(2) length} of the cane brought more insecurities. P8, 13, and 14 claimed that they had difficulties following the cane chassis's exact trajectory as the cane's length caused an offset.
\begin{quote}
P8: "I need to keep remembering the trajectory (of the cane chassis). As there was a distance between the wheel and me, I had not yet walked up to the obstacle when the cane avoided it. If I don't memorize the trajectory and walk strictly according to it, I may run into the obstacle."
\end{quote}

\hspace*{\fill} \

Secondly, there was also a distinction in the \textbf{(3) transmission of grounded kinesthetic feedback}. Despite the cane and car sharing the same type of motors and wheels, P15 and 16 believed the guiding car brought a stronger sense of motion when turning. P16 felt that the inconspicuous kinesthetic feedback reduced his trust in the guiding cane. \textit{P16: "It (cane) has a weaker sense of drive, which made me wonder if it is unreliable."}

\subsubsection{User-related Factors}
Except for on-road scenarios and machine forms, it has been noticed that users' variety can also have an impact on the preference of the autonomy level. P15 was blindfolded because, as a low-vision person, she would like to simulate the feeling of traveling at night. She claimed she wanted to use full autonomy as much as possible because she was very dependent.

With an exactly opposite feeling to P15, P8 complained that it was neither convenient nor safe to be guided by sighted people. She was eager to be in control when traveling and guarantee her own safety. Her first choice of autonomy was partial autonomy. \textit{P8: "I think it is human instinct to subconsciously look out for themselves first, in case of danger. I would be vulnerable as I could only follow him and have no involvement (in making decisions). But when I use partial autonomy, I can take the lead of myself and be more engaged in walking."}

Moreover, another frequently mentioned description of participants' perception of the surrounding area was familiarity. P5, 8, 14, 15, and 17 all agreed that they would use more full autonomy in a familiar environment. \textit{"I am comfortable using full autonomy because I'm well aware of the road condition here," said P14.}

In addition, participants imagined other scenes that they did not encounter during the field study, for instance, taking a stroll in a park and catching a plane or train. P8, 15, and 16 had a consistent view that they would use full autonomy to stroll as it could save effort and use partial autonomy in a hurry as they would trust their control of speed more in critical situations. Distinctively, P17 and 19 chose to use full autonomy in a hurry because they believed the robot could navigate more efficiently. It is worth mentioning that both of them are blind people with little travel frequency.

\section{Discussion}
Findings in control and field study provided a comprehensive view of visually impaired people's perception and utilization of levels of autonomy and machine forms. In summary, users showed a consistent choice for the car-form guiding robot during both controlled and field studies, while the preference for the autonomy level was altered among conditions. Details will be discussed below.

\subsection{Different Levels of Autonomy for BLVs}
The preference for autonomy level can provide a peek at users' mental state. During the controlled study, full autonomy showed a significant advantage in walking efficiency as well as user experience.
Results differ from previous research that higher autonomy can cause lower satisfaction with a sense of control, safety, and trust \cite{zafari2020attitudes, desai2012effects, petersson2011sense}. On the contrary, the full autonomy car-form guiding robots had the highest rating in the controlled environment. 
Feedback from users focused on discussing how intelligent and convenient full autonomy was, to the extent that partial autonomy was considered unnecessary and redundant, even though it outperformed white cane. 

However, during the field study, the environment became complex and uncontrollable, and users' perceptions of the level of autonomy changed as they spent the majority of their travel time using partial autonomy. Full autonomy was supposed as incapable and unreliable when BLV believed the surrounding was dangerous.
Qualitative results showed a connection between the level of autonomy, desired control, trust, and safety, which were uniform with previous research on non-BLV groups \cite{zafari2020attitudes, desai2012effects, petersson2011sense}. Participants believed that a lower level of autonomy could give them more control, which increased their feeling of trust and safety when walking in the real environment.
Even participants who were completely blind and lost 90\% of the channels to perceive information \cite{reuben1988aging, kline1990visibility} believed in themselves more when they assumed the surroundings were dangerous. 
Compared to safety, walking efficiency was not the first priority. Nevertheless, partial autonomy provided a choice to increase mobility greatly and provide psychological and physical safety.

In conclusion, partial autonomy is necessary for situations of uncertainty and complexity and where high walking speeds are required. Moreover, the walking speed controlled by the robot was not comfortable even if participants could adjust the speed by pulling/pushing and pressing the button, which can be explained by the fact that the machine-kind speed control is not identical to manual control \cite{morgado2017but}. Discoveries on partial autonomy can contribute to multiple groups, including disabled and non-disabled. A shared control method offers a way to resolve the paradox of needing intelligence for life-enhancing but lacking trust and safety. Our findings provided empirical evidence for determining autonomy preference among BLV groups in various scenarios.

\subsection{Different Machine Forms for BLVs}
Our studies thoroughly compared BLV people's perceptions of cane and car forms for the machine form. As reviewed previously, the cane was assumed to be a good form for designing guiding robots as it is similar to the traditional white cane. However, it may no longer be the best choice when integrated with mechanical characteristics. Despite the inferiority in portability and familiarity, users still strongly preferred the car-form robot, which could save more effort and provide a higher sense of safety. 

The equal width and proximity to people are important factors for safety, which the cane form can not attain. In the qualitative results, users repeatedly mentioned the car-form robot could provide more sense of safety as they believed the larger width can help to block more obstacles, even though they have been told in advance that the robot could avoid obstacles automatically. 
In fact, the issue that cane-form robots may cause human-environment collisions has been discussed before, and an algorithm to predict human motion and location was proposed to solve this problem when navigating in narrow corridors \cite{wang2021navdog}. This problem can be avoided by altering the machine form as well. 
Participants made suggestions to improve portability while maintaining a high sense of safety, such as retractable length and handrails.

\begin{table}[h]
    \caption{Machine Form Specifics}
    \label{form}
    \begin{tabular}{ccccc}
    \toprule
     & \multicolumn{2}{c}{Distance from Human (m)} & \multicolumn{2}{c}{Steering Force (N)} \\
     & Steering wheel & Forward wheel & To wheel & To hand \\ 
     \midrule
    Cane & 0.98 & 0.70 & 7.14 & 2.86 \\
    Car & 0.71 & 0.16 & 5.00 & 17.20 \\
    \bottomrule
    \end{tabular}
\end{table}

Moreover, table \ref{form} demonstrated the form specifics of the robot cane and car. It is noticeable that the distance between the steering wheel of the guiding cane and the users is greater than that of the guiding car, confirming participants' perception in the qualitative result that it was difficult to follow the cane robot's motion.

Participants mentioned in the interview that they believed the steering force of the cane-form robot was weaker than the car-form robot, which provided them with a feeling of uncertainty.
The steering force in table \ref{form} provides evidence for the gap in kinesthetic feedback of cane and car form guiding robots. Firstly, because of the lack of auxiliary wheels, the steering wheel of the guiding cane would provide greater support, therefore greatly steering friction. However, as the rotation centers of the cane and car are both the forward wheel, the rotation radius of the cane is significantly smaller. According to the definition equation of moment, ${\tau = F_{1}r_{1} = F_{2}r_{2}}$, the force acting on the users' hands of the cane that indicates turning is smaller than the car, which results in weaker transmission of the kinesthesia. Turning is essential to guiding, so the weak transmission of movement can make users feel uncertain, resulting in a low sense of trust and safety. Moreover, the structure of the stable car form is more in line with robot design principles that can carry more components and is simple to develop.

Compared with traditional white canes and guide dogs, our robots are more intelligent and safer. As shown in the controlled study, the guiding robots expressed better navigation efficiency and less mental workload than the baseline, which can increase BLV groups' mobility independence and well-being.

Previous studies have tried cane and car forms as the guiding robot design structures. Some researchers were inspired by the familiar and lightweight canes used by visually impaired people \cite{chuang2018deep, varela2020robotic}, others believed the form of cars were safer, closer to a guide dog, and easier to blend in with the crowd to avoid embarrassment \cite{guerreiro2019cabot, tobita2018structure}. In this study, we contributed to a cross-sectional comparison between cane and car form structures. Both quantitative and qualitative results revealed that even though the BLV users were more experienced in using white canes, the form of a car was more appropriate and preferred as a guiding robot.

\subsection{Design Implications for Assistive Robots}
To explore whether visually impaired people require a degree of control when navigating with an automated robot, we developed two autonomy-switchable guiding robots with high usability that have not been attempted before. The robots were designed to be robust to discover users' perceptions and performance in the field. They had a reasonable user interface and logic for autonomy switching, which enabled participants choices between partial and full autonomy under different circumstances. 

Participants switched to partial autonomy because they needed more dominance and changed to full autonomy to save effort and help them maintain a higher walking speed. As neither of these states would last the entire journey, it was essential to have the ability to transition between them. 

For the community of disabled people, partially automated assistive robots allow for more control and flexibility for the user, who can still perform some tasks independently while the robot assists. Additionally, autonomy-switchable assistive robots can be better suited to handle the unpredictable nature of many disabilities by adjusting to changes and providing the right amount of assistance as needed. Moreover, partial autonomy can help reduce the risk of injury, as full autonomy may not react to changes in the environment or user's condition in the same way that a human can, which could potentially lead to accidents or injuries.

In conclusion, the autonomy-switchable guiding cane and car with high usability ensured the explorations on levels of autonomy and machine forms can be conducted in various environments. Our findings on levels of autonomy are the first to understand the need for a sense of control, trust, and safety in guiding robots for the BLV group, which can help to solve psychological issues caused by full autonomy assistive robots and the performance issues caused by partial autonomy and manual assistive methods mentioned by previous studies \cite{holbo2013safe, kim2011autonomy, bhattacharjee2020more}. The evidence on the advantages and disadvantages of the cane-form and car-form guiding robots can provide instruction for designers. The discovery of the machine forms provides the first cross-sectional comparison of guiding robots design by the HCI accessibility community.

\subsection{Limitation and Future Work}
There were two limitations in this research. One concern was that our results were somewhat influenced by the design of these two specific robots. In the current version, the SUS score of the cane was 81, lower than that of the car, 88. The final results did show a preference for the car form, but if another more usable cane were substituted, it could be reasonably inferred that the results would be biased.
In addition, there were certain disadvantages to the structure and performance of the robot design. The space between the bottom of the car and the participant's legs was so cramped and limited that it would bring uncomfortable feelings. 
Also, the speed was still considered "slower than expected" in full autonomy. It was observed that users would subconsciously slow down before performing operations (pressing the button) when switching from partial to full autonomy. Although these shortcomings affected the experience to a certain extent, they were not used as a variable in exploring the preference for machine form and autonomy, so they did not substantially impact the final conclusion.
In the future, it is worthwhile to optimize robot design by adding an inverted L-shaped handle in the car and adjusting the interaction of switching autonomy. Additionally, robot-controlled speed is not as natural as manually-controlled speed in guiding assistive robots, so further research should be undertaken to apply the logic and pattern of manually controlled speed to robot control to achieve the expected comfort experience.
 
Another concern lay in the scarcity of BLV participants, which could be detrimental to generalization in a demographic sense. The BLV community was jointly composed of many kinds of blind people, including guide dog users, elderly blind people, etc. Due to the small proportion of such groups in the BLV community, they were not targeted when calling for participants. In addition, the online recruitment method also determined the general portrait of these participants, who had stable jobs, friend circles, mobile phones, educational backgrounds, relatively rich social activities, and lived in cities. Therefore, the participants' demographics were limited, so We need more samples and evidence.
Future work should focus on supplementing more BLV participant data to consolidate the general conclusion. Moreover, it is recommended to explore how demographic indicators of participants, such as age groups, levels and duration of BLV, and travel aids affect their preference for machine form and autonomy. 

\section{Conclusion}
In this research, a controlled and a field study were conducted to discover the level of autonomy and machine form's impact on BLV's perception and utilization based on two autonomy-switchable guiding robots with a cane and a car form presented.
People assumed they had lost most of the sensory channels to perceive the surrounding environment and that they could not be the "follower" of the guiding robot and the "boss" to control simultaneously. However, conclusions have been drawn for autonomy preference that people with vision loss still have desires and requirements in control when performing a navigation task with guiding robots.
With increased satisfaction with the sense of control, users will feel more trust and safety, providing a better experience. Partial autonomy and full autonomy are both necessary when developing a guiding robot. 
As for the machine form that carries navigation functions, the car form can be a better choice than imitating the shape of a white cane, which can provide more sense of safety and be better compliant with robot design principles.
The guiding robots used in this research were a preliminary exploration. Future improvements can be made to a more intelligent shared control algorithm and a more fully-functional navigation system. Findings on the level of autonomy and machine form may provide constructive instruction on developing more user-friendly and human-centered assistive and collaborative robots for all humankind.

\begin{acks}
This project is supported by National Natural Science Foundation Youth Fund 62202267.
We want to thank all the participants in this study.
\end{acks}

\bibliographystyle{ACM-Reference-Format}
\bibliography{Reference.bib}

\newpage
\appendix
\setcounter{figure}{0}
\renewcommand\thefigure{\Alph{section}\arabic{figure}}
\setcounter{table}{0}
\renewcommand\thetable{\Alph{section}\arabic{table}}  

\section{Appendix: Controlled study trail-following method}
\begin{figure}[h]
    \centering
    \includegraphics[width=0.5\textwidth]{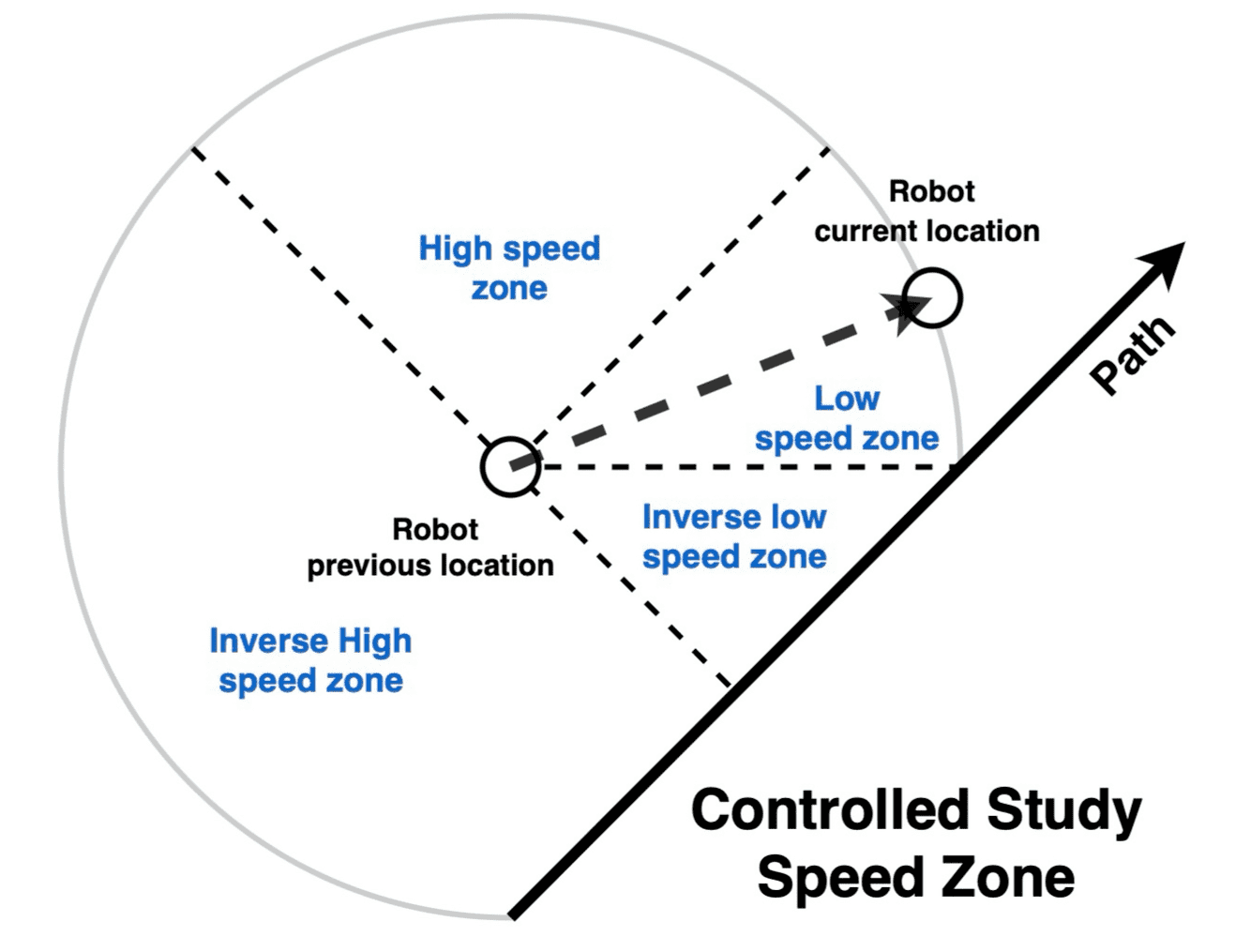}
    \caption{A demonstration of the robot trail-following method in the controlled study. Each speed zone was responsible for robot steering instruction.}
    \label{navigation}
\end{figure}

With the HTC VIVE positioning system, the server laptop could update the robot's location at a constant frequency. After each update, it calculated the shortest distance between the robot and the pre-set route, the direction of the current path section, and the real-time heading angle to give a steering pulse instruction for the motor to get back on track. We calculated the guiding robot's forward angle and path angle and set up three zones by comparing them (see figure \ref{navigation}), where each zone had its steering pulse instruction. We defined a constant length steering pulse by sending steering speed first (which also means start steering) followed by a time delay and ending it with a stop steering instruction. The robot would receive a low-speed steering pulse in the low-speed zone, making it approach the path gently. Similarly, a higher-speed steering pulse was sent to quickly make the robot back on route in the high-speed zone. In particular, the server PC would send a steering pulse in the opposite turning direction in the inverse-speed zone to prevent overturning. Furthermore, it is worth mentioning that the steering speed in each zone can automatically adapt to the machine form and the forward velocity, regardless of whether it is in partial or full autonomy. 
\label{Controlled study trail-following method}

\newpage
\section{Appendix: Controlled Study Interview}
\label{control_interview}
\begin{itemize}
    \item What is your preference for the following guiding assistance: white cane, normal cart, robot cane with partial autonomy, robot cane with full autonomy, robot car with partial autonomy, and robot car with full autonomy?
    \item What are the reasons for your least favorite one?
    \item What are the reasons for your favorite one?
    \item Do you have any differences in feelings about the cane and car form?
    \item Do you have any differences in feelings about full autonomy, partial autonomy, and baseline? 
    \item Is there a difference in the degree of control you feel when using these robots? Which robot do you like the most to feel in control? Why?
    \item Would there be a sense of embarrassment if you use these robots in daily life?
    \item Apart from the above questions, are there any other differences in the way you feel when using guiding robots with different levels of autonomy and machine forms?
\end{itemize}

\section{Appendix: Field Study Interview}
\label{field_interview}
\begin{itemize}
    \item Review participants' behaviors during the travel and ask follow-up questions.
    \item Do you have any differences in feelings about full autonomy and partial autonomy? (On the aspect of satisfaction, safety, trust, embarrassment, control, etc)
    \item Do you have any differences in feelings about cane form and car form? (On the aspect of satisfaction, safety, trust, embarrassment, control, etc)
    \item What factor affected your speed? Under what conditions would you accelerate and accelerate? Would you walk faster in a safe environment?
    \item What factor affected your choice of autonomy level? In what scenarios would you prefer to use full autonomy? In what scenarios would you prefer to use partial autonomy? (Objectively and subjectively)
    \item In what scenarios would you prefer a cane-form robot? In what scenarios would you prefer a car-form robot?
    \item Apart from the above questions, are there any other differences in the way you feel when using guiding robots with different levels of autonomy and machine forms?
\end{itemize}
\section{Appendix: Scales}
This section contains the PIADS and SUS scales that were used in the controlled study.

\label{appendix: scales}
\begin{table*}[t]
\caption{Modified Psychosocial Impact of Assistive Devices Scale (PIADS)}
\begin{tabular}{@{}llclllllll@{}}
\toprule
   &                                                    &                                                                                     & -3 & -2 & -1 & 0 & 1 & 2 & 3 \\ \midrule
1  & Competence                             &\begin{tabular}{ccc} Ability to do well the important things\\ you need to do in life\end{tabular}                      &    &    &    &   &   &   &   \\
2  & Happiness                                          &\begin{tabular}{ccc} Gladness, pleasure; satisfaction with life\end{tabular}                                          &    &    &    &   &   &   &   \\
3  & Independence                                       & \begin{tabular}{ccc}Not dependent on, or not always needing \\help from, someone or something\end{tabular}&    &    &    &   &   &   &   \\
4  & Adequacy                                           &\begin{tabular}{ccc} Capable of handling life situations, and \\handling little crises\end{tabular}                     &    &    &    &   &   &   &   \\
5  & Confusion                                          &\begin{tabular}{ccc} Unable to think clearly, act decisively\end{tabular}                                             &    &    &    &   &   &   &   \\
6  & Efficiency                                         &\begin{tabular}{ccc} Effective management of day to day tasks\end{tabular}                                            &    &    &    &   &   &   &   \\
7  & Self-esteem                                        &\begin{tabular}{ccc} How you feel about yourself, and like \\yourself as a person\end{tabular}                          &    &    &    &   &   &   &   \\
8  & Productivity                                       &\begin{tabular}{ccc} Able to get more things done in a day\end{tabular}                                               &    &    &    &   &   &   &   \\
9  & Security                                           &\begin{tabular}{ccc} Feeling safe rather than feeling vulnerable \\or insecure\end{tabular}                             &    &    &    &   &   &   &   \\
10 & Frustration                                        & \begin{tabular}{ccc}Being upset about lack of progress in achie\\-ving your desires; feeling disappointed   \end{tabular}&   &    &    &   &   &   &   \\
11 & Usefulness                                         &\begin{tabular}{ccc} Helpful to yourself and others; can get\\ things done\end{tabular}                                 &    &    &    &   &   &   &   \\
12 & Self-confidence                                    &\begin{tabular}{ccc} Self-reliance; trust in yourself and your abilities\end{tabular}                                 &    &    &    &   &   &   &   \\
13 & Expertise                                          &\begin{tabular}{ccc} Knowledge in a particular area or occupation\end{tabular}                                        &    &    &    &   &   &   &   \\
14 & Skillfulness                                         & \begin{tabular}{ccc}  Able to show your expertise; perform tasks well \end{tabular}                                     &    &    &    &   &   &   &   \\
15 & Well-being                                         & \begin{tabular}{ccc}       Feeling well;\\ optimistic about your life and future  \end{tabular}                               &    &    &    &   &   &   &   \\
16 & Capability                                         & Feeling more capable; able to cope                                                  &    &    &    &   &   &   &   \\
17 & Quality of life                                    & How good your life is                                                               &    &    &    &   &   &   &   \\
18 & Performance                                        & Able to demonstrate your skills                                                     &    &    &    &   &   &   &   \\
19 & Sense of power                                     &\begin{tabular}{ccc} Sense of inner strength; feeling that you have \\significant influence over your life \end{tabular}&    &    &    &   &   &   &   \\
20 & Sense of control                                   &\begin{tabular}{ccc} Sense of being able to do what you want in \\your environment\end{tabular}                         &    &    &    &   &   &   &   \\
21 &\begin{tabular}[c]{@{}l@{}} Satisfaction with sense\\ of control\end{tabular}                 & Satisfaction with this sense of control you rated above                                             &    &    &    &   &   &   &   \\
22 & Embarrassment                                       & Feeling awkward or ashamed                                                          &    &    &    &   &   &   &   \\
23 & Trust                                              & Feeling trust in assistive devices                                                 &    &    &    &   &   &   &   \\
24 & Willingness to take chances                        &\begin{tabular}{ccc} Willing to take some risks; willing to take on \\new challenges\end{tabular}                       &    &    &    &   &   &   &   \\
25 & Ability to participate                             & Ability to join in activities with other people                                     &    &    &    &   &   &   &   \\
26 & Eagerness to try new things                        &\begin{tabular}{ccc} Feeling adventuresome and open to \\new experiences\end{tabular}                                   &    &    &    &   &   &   &   \\
27 &\begin{tabular}{ccc} Ability to adapt to the \\activities of daily living \end{tabular}& \begin{tabular}{ccc}Ability to cope with change; ability to make \\basic tasks more manageable            \end{tabular}&    &    &    &   &   &   &   \\
28 &\begin{tabular}{ccc} Ability to take advantage\\ of opportunities\end{tabular}          & \begin{tabular}{ccc}Ability to act quickly and confidently when \\there is a chance to improve something  &\end{tabular}\    &    &    &   &   &   &   \\
29 &\begin{tabular}{ccc} Rate the confidence of \\your answer(1-5)\end{tabular}            &                                                                                     &    &    &    &   &   &   &   \\ \bottomrule
\end{tabular}

\end{table*}

\begin{table*}[]
\caption{System Usability Scale (SUS)}
\begin{tabular}{@{}|l|l|l|l|l|l|@{}}
\toprule
                                                                                             & 1 & 2 & 3 & 4 & 5 \\ \midrule
1. I think that I would like to use this system frequently                                   &   &   &   &   &   \\ \midrule
2. I found the system unnecessarily complex                                                  &   &   &   &   &   \\ \midrule
3. I thought the system was easy to use                                                      &   &   &   &   &   \\ \midrule
4. I think that I would need the support of a technical person to be able to use this system &   &   &   &   &   \\ \midrule
5. I found the various functions in this system were well integrated                         &   &   &   &   &   \\ \midrule
6. I thought there was too much inconsistency in this system                                 &   &   &   &   &   \\ \midrule
7. I would imagine that most people would learn to use this system very quickly              &   &   &   &   &   \\ \midrule
8. I found the system very cumbersome to use                                                 &   &   &   &   &   \\ \midrule
9. I felt very confident using the system                                                    &   &   &   &   &   \\ \midrule
10. I needed to learn a lot of things before I could get going with this system              &   &   &   &   &   \\ \bottomrule
\end{tabular}
\end{table*}

\end{document}